\documentclass[aps,prl,twocolumn,showpacs,floatfix,nobibnotes,superscriptaddress,longbibliography]{revtex4-1}

\usepackage{amsmath}
\usepackage{inputenc}
\usepackage{amssymb}
\usepackage{graphicx}
\usepackage{verbatim}
\usepackage{epsfig}
\usepackage{bm}
\usepackage{color}
\usepackage{float}
\usepackage{dcolumn}
\usepackage{multirow} 
\usepackage[unicode=true,
  linktocpage,
  linkbordercolor={0.5 0.5 1},
  citebordercolor={0.5 1 0.5},
  colorlinks=true,
  linkcolor=blue,
  citecolor=blue,
  urlcolor=blue]{hyperref} 
\usepackage{cgp_macros}
\usepackage{lipsum} 
\usepackage{placeins} 

\usepackage[normalem]{ulem}

\graphicspath{{.}}

\newcommand{\beq}{\begin{equation}}
\newcommand{\eeq}{\end{equation}}
\newcommand{\beqa}{\begin{eqnarray}}
\newcommand{\eeqa}{\end{eqnarray}}

\newcommand{\bra}[1]{\left\langle #1 \right|}
\newcommand{\ket}[1]{\left| #1 \right\rangle}
\newcommand{\braket}[2]{\left\langle #1 \vert #2 \right\rangle}

\newcommand{\NNLOsat}{NNLO$_{\rm sat}$}
\newcommand{\NNLOgod}{$\Delta$NNLO$_{\rm GO}$(450)}

\begin{document}

\title{\emph{Ab initio} computation of the  longitudinal response function in $^{40}$Ca}

\author{J.~E.~Sobczyk}
\affiliation{Institut f\"ur Kernphysik and PRISMA$^+$ Cluster of Excellence, Johannes Gutenberg-Universit\"at, 55128
  Mainz, Germany}

\author{B.~Acharya}
\affiliation{Institut f\"ur Kernphysik and PRISMA$^+$ Cluster of Excellence, Johannes Gutenberg-Universit\"at, 55128
  Mainz, Germany}
  
\author{S.~Bacca}
\affiliation{Institut f\"ur Kernphysik and PRISMA$^+$ Cluster of Excellence, Johannes Gutenberg-Universit\"at, 55128
  Mainz, Germany}
\affiliation{Helmholtz-Institut Mainz, Johannes Gutenberg-Universit\"at Mainz, D-55099 Mainz, Germany}
  
\author{G.~Hagen}

\affiliation{Physics Division, Oak Ridge National Laboratory,
Oak Ridge, TN 37831, USA} 
\affiliation{Department of Physics and Astronomy, University of Tennessee,
Knoxville, TN 37996, USA}

\begin{abstract}
 We present a consistent \emph{ab initio} computation of the longitudinal response function $R_L$ in $^{40}$Ca using the coupled-cluster and Lorentz integral transform methods starting from chiral nucleon-nucleon and three-nucleon interactions. We validate our approach by comparing our results for $R_L$ in $^4$He and the Coulomb sum rule in $^{40}$Ca against experimental data and other calculations. For $R_L$ in $^{40}$Ca we obtain a very good agreement with experiment in the quasi-elastic peak up to intermediate momentum transfers, and we find that final state interactions are essential for an accurate description of the data. This work presents a milestone towards \emph{ab initio} computations of neutrino-nucleus cross sections relevant for experimental long-baseline neutrino programs.
\end{abstract}

\maketitle 

Understanding a wide variety of nuclear phenomena in terms of constituent nucleons is a major ongoing initiative in nuclear theory~\cite{Heiko}.  Theoretical predictions that start from the forces among nucleons and their interactions with external probes as described by chiral effective field theory are arguably the doorway to connect experimental observations with the underlying fundamental theory of quantum chromo-dynamics~\cite{vankolck1994,bedaque2002,epelbaum2009,machleidt2011}.  This approach is key to interpret existing data, provide guidance for future experiments, and support interdisciplinary efforts at the interface with nuclear physics, such as neutrino physics~\cite{Nustec}.

Neutrino oscillation experiments aim at addressing some of the biggest unanswered questions in physics by measuring the charge conjugation-parity  violating phase in the lepton sector of the Standard Model of particle physics. For the current neutrino oscillation experiments the systematic errors are at the order of $\sim 10\%$ and largely influenced by considerable cross-section uncertainties. Next generation experiments set their precision goal much higher. The T2HK~\cite{hyperk} and DUNE~\cite{DUNE} experiments aim at achieving much smaller statistical fluctuations, comparable with present systematic errors. It is therefore crucial to control the systematics, whose major part comes from the limited precision of theoretical modeling of neutrino-nucleus cross sections. Furthermore, the exposure needed to achieve a desired sensitivity  also depends on the ability of reducing systematic errors.
The models which are presently in use, particularly the ones implemented in the Monte Carlo event generators, should be benchmarked with the predictions given by \emph{ab initio} models of nuclear dynamics for relevant nuclei such as $^{12}$C, $^{16}$O and $^{40}$Ar.

Due to recent developments of accurate quantum many-body methods with controlled approximations, ever-increasing computing power, and advancements in 
the description of nuclear interactions and electroweak currents, we are now entering an era where the \emph{ab initio} description of lepton-nucleus scattering is becoming possible. The Green's Function Monte Carlo (GFMC) method was used to calculate  nuclear responses of $^4$He and $^{12}$C~\cite{Lovato2014,Lovato:2015qka}, and was recently able to make direct comparisons with the neutrino-nucleus experimental cross sections~\cite{Lovato:2017cux,lovato2020}. Using the same dynamical ingredients, other simplified methods are being developed to reduce the computational load and address the quasi-elastic peak~\cite{STA}.
In another set of studies, the lepton-nucleus scattering cross sections of $^4$He, $^{16}$O and $^{40}$Ar were obtained using spectral functions calculated in the  self-consistent Green's function method with final-state interactions included using mean-field potentials~\cite{Rocco:2018vbf, Barbieri:2019ual}.

 In this Letter, we lay out the tools for an \emph{ab initio} method that accurately accounts for final state interactions, consistently with the treatment of initial state interactions, and demonstrate its advantages by comparing to available longitudinal electron scattering data for $^{40}$Ca. We base our approach on the coupled--cluster (CC) method~\cite{coester1958,coester1960,kuemmel1978,mihaila2000b,dean2004,wloch2005,hagen2008,hagen2010b,binder2013b,hagen2014}, which stands out as one of the most suitable and promising methods for calculations involving medium-mass and heavy nuclei due to the polynomial scaling of its computational cost with the mass number $A$. Initially applied to closed-shell nuclei (see Ref.~\cite{hagen2014} for a review), it has since been extended to accurately describe doubly open-shell neighbors such as $^{40}$Ar~\cite{liu2019,Payne:2019wvy}, and more recently starting from an axially deformed reference state entire isotope chains~\cite{novario2020,koszorus2021}. Combining CC with the Lorentz integral transform method~\cite{efros1994,efros2007}, the LIT-CC approach extends the reach of this theory to processes involving excitation of bound nuclear states to the continuum. Originally applied to calculate low-energy nuclear dipole responses~\cite{bacca2013,bacca2014},  recently it was extended to compute the Coulomb sum rule for $^4$He and $^{16}$O~\cite{Sobczyk:2020qtw}. By devising a method to project out the spurious center-of-mass (CoM) excitations, Ref.~\cite{Sobczyk:2020qtw} has also tackled the major technical challenge of removing CoM contaminations in calculations utilizing translationally non-invariant nuclear electroweak operators. 
These developments open the door to go beyond the sum rule calculations and gain deeper insights into the dynamics of the nucleus
by computing the nuclear response functions.
With the goal of eventually applying the theory to neutrino-nucleus scattering, where experimental data are scarce or imprecise, we first benchmark our results for inelastic  electron-scattering by comparing them with existing data
for $^{40}$Ca.

The inclusive cross section of this process can be expressed in terms of two response functions: the longitudinal, $R_L(\omega, q)$, and the transverse, $R_T(\omega, q)$, where $\omega$ is the energy transferred from the electron to the nucleus. 
These are induced by the charge and the current operator, respectively, and can be experimentally disentangled using the so-called Rosenbluth separation. We study the longitudinal response in this work and defer the transverse response, which receives large two-nucleon electromagnetic current contributions~\cite{lovato2020}, to a future work. Formally, the longitudinal response function can be defined as 
\beq
\label{frisp}
R_L(\omega,q)\!=\!\int \!\!\!\!\!\!\!\sum _{f} 
|\!\left\langle \Psi_{f}| \rho(q)| 
\Psi _{0}\right\rangle\!|
^{2}\,\delta\!\!\left(\!E_{f}+\frac{q^2}{2M}-E_{0}-\omega \! \right), 
\eeq
where $M$ is the mass of the target nucleus, and $| \Psi_{0/f} \rangle$ and $E_{0/f}$ respectively denote the  initial/final-state nuclear wave functions and energies, which we compute using nucleon-nucleon and three-nucleon forces from chiral effective field theory. In order to estimate the sensitivity of our results on the employed Hamiltonian we use two different chiral interactions, namely \NNLOsat\ \cite{Ekstrom:2015rta} and \NNLOgod\ \cite{jiang2020}. These interactions are both given at next-to-next-to-leading order in the chiral expansion and employ a regulator cutoff of $450$~$\mathrm{MeV}/c$, but they differ in that \NNLOgod\ includes explicit $\Delta$-isobars in its construction while \NNLOsat\ does not.  These interactions are well suited for our study  of $^{4}$He and $^{40}$Ca as they have been shown to provide an accurate description of radii and binding energies of light and medium-mass nuclei nuclei, and the saturation point of symmetric nuclear matter~\cite{Ekstrom:2015rta,jiang2020}.

The charge density operator considered in this work is 
\beq
{\rho}(q)= \frac{e}{2} \sum_{i=1}^A \,\left(G_E^S(Q^2) + \tau_i^3 \, G_E^V(Q^2)\right)  \exp{(i {\bf q} \cdot {\bf r}_i)} \,, 
\label{eq:operator}
\eeq
where $e$ is the proton charge, while ${\bf r}_i$ and $\tau_i^3$ are the coordinate and the third isospin component of nucleon~$i$. We use the parametrization of Ref.~\cite{Kelly:2004hm} for the nucleon isoscalar/isovector electric form factors, $G_E^{S/V}(Q^2)$. The Darwin-Foldy and the spin-orbit relativistic corrections, as well as the two-nucleon current contributions, are not included in Eq.~\eqref{eq:operator} since we strive for consistency between the power-counting and truncation in the chiral expansions of the current and the interactions. Specifically, corrections to Eq.~\eqref{eq:operator} are at least four orders higher in the chiral expansion when the inverse of the nucleon mass is counted as two chiral orders~\cite{Krebs:2019aka}, which is beyond the order at which the interactions we use are truncated.

The sum over $\Psi_f$ in Eq.~\eqref{frisp} poses a serious computational challenge, since it involves an integration over the continuum states, when $\omega$ is above the particle emission threshold $\omega_{th}$. 
To overcome this issue, we use the LIT method, where through the application of a Lorentzian-kernel transform
\begin{equation}
{\cal L}_L(\sigma,q)=\frac{\sigma_I}{\pi}\int d\omega
\frac{R_L(\omega,q)}
{(\omega-\sigma_R)^{2}+\sigma_I^{2}}
= \langle\widetilde{\Psi}_{\sigma,q}^{\rho}
|\widetilde{\Psi}_{\sigma,q}^{\rho}\rangle  \label{lorentz_transform} 
\end{equation}
with $\sigma_I\ne0$,
 one reduces the problem to solving
\begin{equation}
(H-E_{0}-\sigma)|\widetilde{\Psi}_{\sigma,q}^{\rho}
\rangle={\rho}(q)|{\Psi_{0}}\rangle\,,\label{liteq}
\end{equation}
where $H$ denotes the nuclear Hamiltonian. 
Effectively, $\widetilde{\Psi}_{\sigma,q}^{\rho}$ is the solution of a bound-state ``Schr\"{o}dinger-like'' equation with a source term, which can be solved also in coupled-cluster theory.

The CC method allows for the inclusion of many-body correlations as a controlled expansion by writing the nuclear wave function as $\vert \Psi\rangle = e^T\vert \Phi_0\rangle$. Here $\vert \Phi_0 \rangle$ is a suitably chosen reference state, and $T = T_1 + T_2+ \ldots$ is a linear expansion in particle-hole excitations typically truncated at some low excitation rank. In this work we truncate $T = T_1 + T_2 $ which is known as the coupled-cluster singles and doubles (CCSD) method. Inserting the CCSD wave function into the many-body Schr\"odinger equation and projecting from the left with $e^{-T}$, it is seen that the reference state $\vert \Phi_0 \rangle$ is the ground-state of the similarity transformed normal-ordered Hamiltonian $\overline{H}_N= e^{-T} H_N e^T$. 
In the LIT-CC formulation one has to employ
  the equation-of-motion coupled-cluster technique (EOM-CC)~ \cite{stanton1993}  
with a  source term (see r.h.s.~of  Eq.~(\ref{liteq})) and the similarity 
transformed  normal-ordered operator
 $\overline{\Theta}_N \equiv e^{-T}\Theta_Ne^T$~\cite{miorelli2018}.
Here, $\Theta$ are the rank-$J$ multipoles of the electromagnetic charge operator given by Eq.~\eqref{eq:operator}. To obtain the LIT, we perform EOM-CC calculations for each multipole $[\rho(q)]^J$, and perform the sum over all multipoles at the end~(see also Ref.~\cite{bijaya}).

The response function $R_L(\omega,q)$ for a given value of $q$ is then obtained by inverting
the integral transform from Eq.~\eqref{lorentz_transform}. 
To perform the inversions, which require the solution of an ill-posed problem,  we perform the expansion 
$ R_L(\omega)=
\sum_i^N c_i \omega^{n_0}  e^{-\frac{\omega}{\beta i}}$
and seek for stable solutions by varying the non-linear parameter $\beta$ (as well as $n_0$) in a certain range.
The inversion procedure involves the determination of the coefficients $c_i$  of the $N$ basis functions by a least-squares fit~\cite{efros2007}. We impose $R(\omega)$ to be zero for $\omega \le \omega_{th}$, using the values we obtain for a given nuclear Hamiltonian in the CCSD approximation.
We estimate the uncertainty associated with the inversion procedure by inverting LITs with three different values of $\sigma_I=5,10$ and 20~MeV and by varying $N$ from 6 to 9.

In all our results we employ a model space consisting of 15 major oscillator shells ($e_{\rm max} = 2n+l = 14$) with an additional cut on the matrix elements of the three-nucleon force given by $e_{3{\rm max}} =   2n_1 + l_1  + 2n_2 + l_2 + 2n_3 + l_3  \le 16$. We checked that we can reach a satisfactory convergence of $\mathcal{L}_L$ in terms of the single-particle model space size $e_{\rm max}$. The latter can be tested, e.g., by studying the residual dependence on the underlying harmonic oscillator frequency $\hbar \Omega$. In particular,
for LITs with $\sigma=20$ MeV we estimate the convergence in the quasi-elastic peak to be at the 2$\%$ level for $q\le 350$ MeV/c and of $4\%$  for $q\ge 400$ MeV/c, by varying $\hbar\Omega $ in the range 18  to 22~MeV.

\begin{figure}[hbt]
  \includegraphics[width=0.4\textwidth]{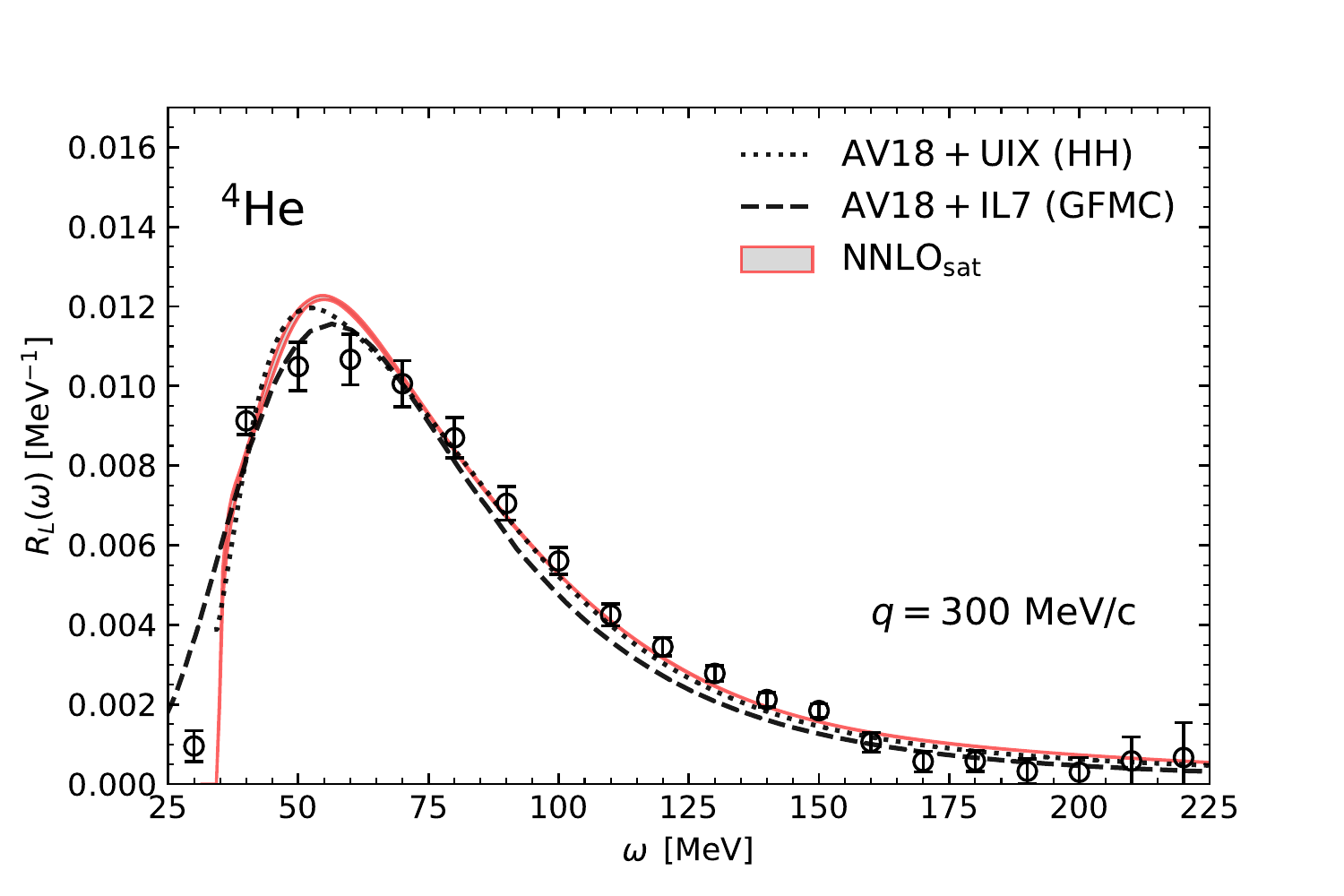}
  \caption{Longitudinal response function for $^4$He at $q=300$ MeV/c. HH results taken from Ref.~\cite{barnea2001}, GFMC results from Ref.~\cite{Rocco:2018tes}, and experimental data from Ref.~\cite{world_data}. }
  \label{bench}
\end{figure}

{\it Benchmark on the $^4\mathrm{He}$ nucleus---} We begin by presenting our results for $R_L$ in the case of $^4$He at $q=300$ MeV/c. In Fig.~\ref{bench}, we show calculations performed with the  \NNLOsat\  interaction in the CCSD scheme for an underlying harmonic oscillator frequency of $\hbar \Omega = 16$ MeV. 
Here  the small band reflects only the uncertainty associated with the LIT inversion.
For comparison, we also show calculations performed with the hyperspherical harmonics method (HH)~\cite{BaccaPRL2009} using the AV18+UIX potential and Green Function Monte Carlo (GFMC)~\cite{Rocco:2018tes} calculations that used the AV18+IL7 potential. 
We obtain very good agreement with the experimental data as well as with other theoretical calculations. This comparison corroborates our method and further validates the protocol we developed in Ref.~\cite{Sobczyk:2020qtw} to remove center of mass contamination.

\begin{figure}[hbt]
	\includegraphics[width=0.4\textwidth]{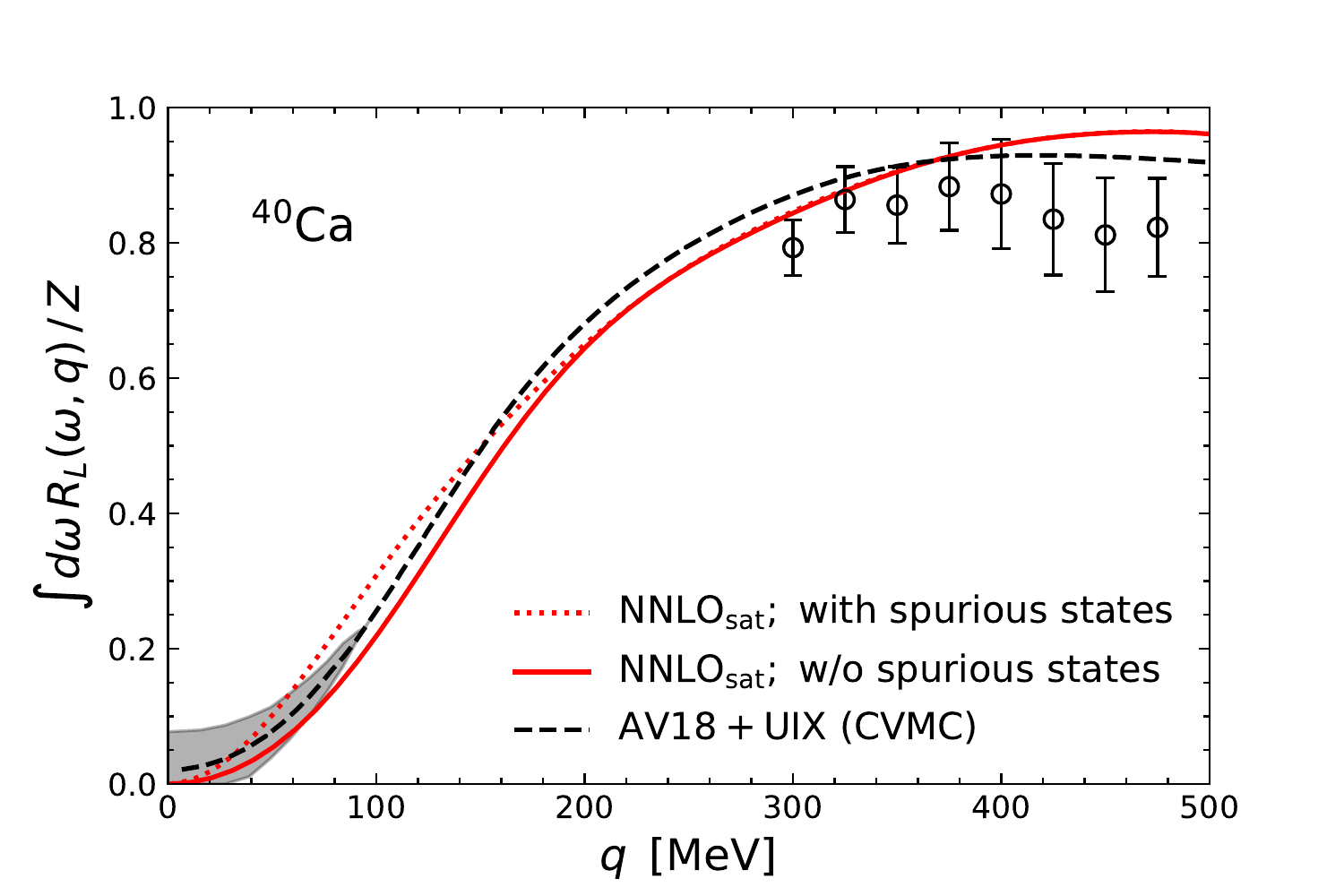}
	\caption{$^{40}$Ca results for Coulomb sum rule for N2LO$_{\mathrm{sat}}$  and $\hbar \omega=22$ MeV compared with CVMC results of Ref.~\cite{Lonardoni} and experimental data taken from Ref.~\cite{Williamson:1997zz}.}
	\label{CSR}
\end{figure}

\begin{figure*}[thb]
	\includegraphics[width=0.4\textwidth]{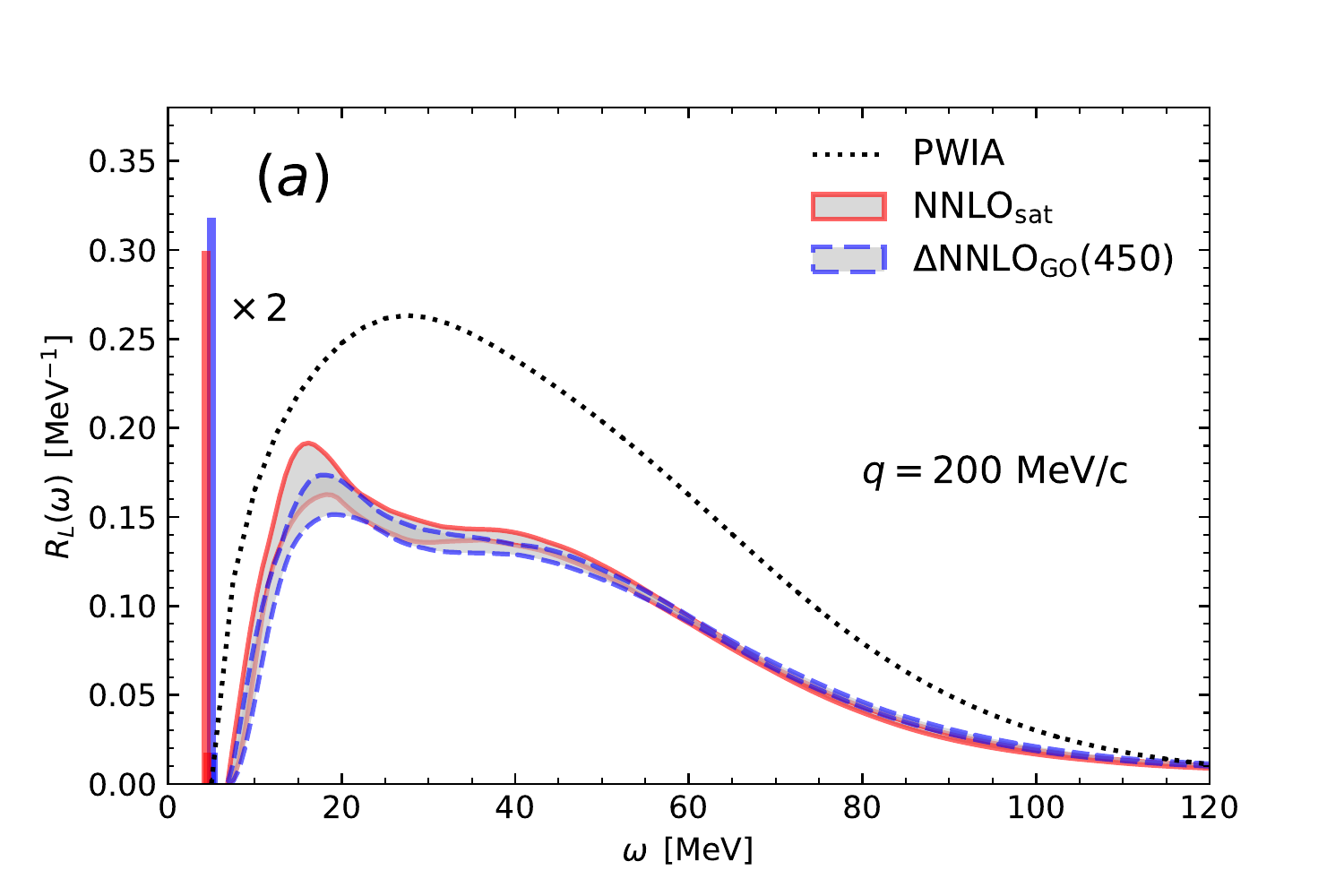}
	\includegraphics[width=0.4\textwidth]{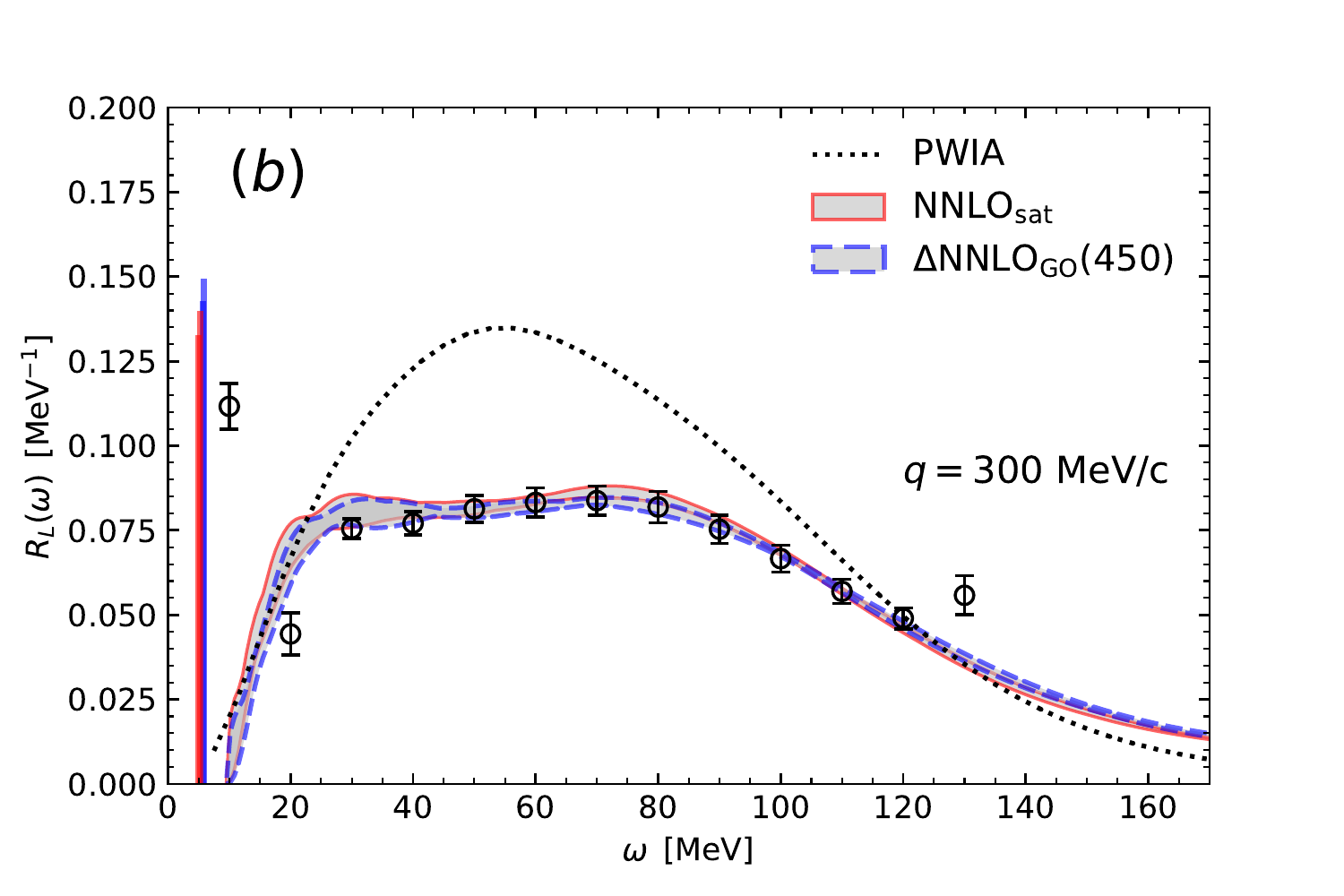}
	\includegraphics[width=0.4\textwidth]{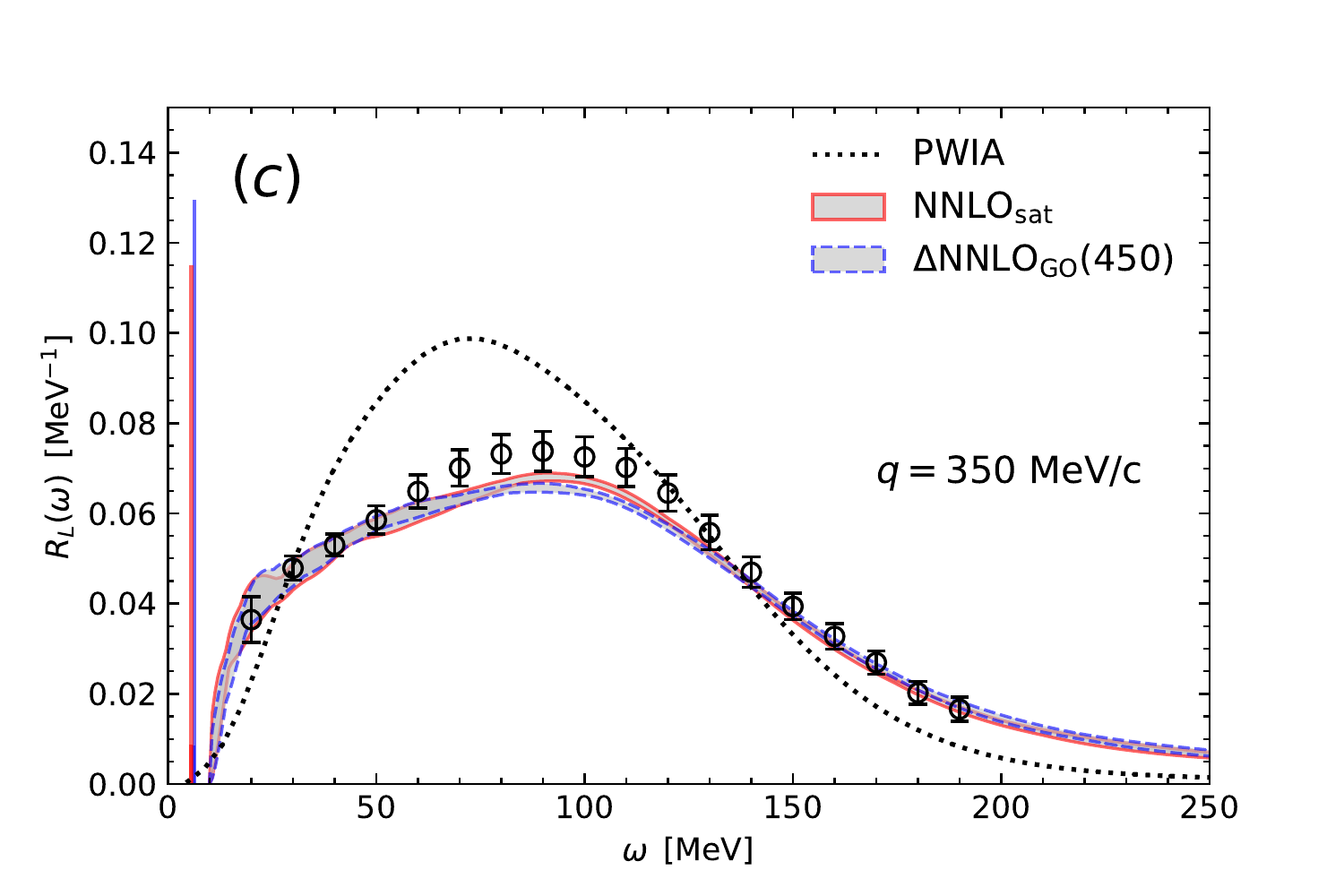}
	\includegraphics[width=0.4\textwidth]{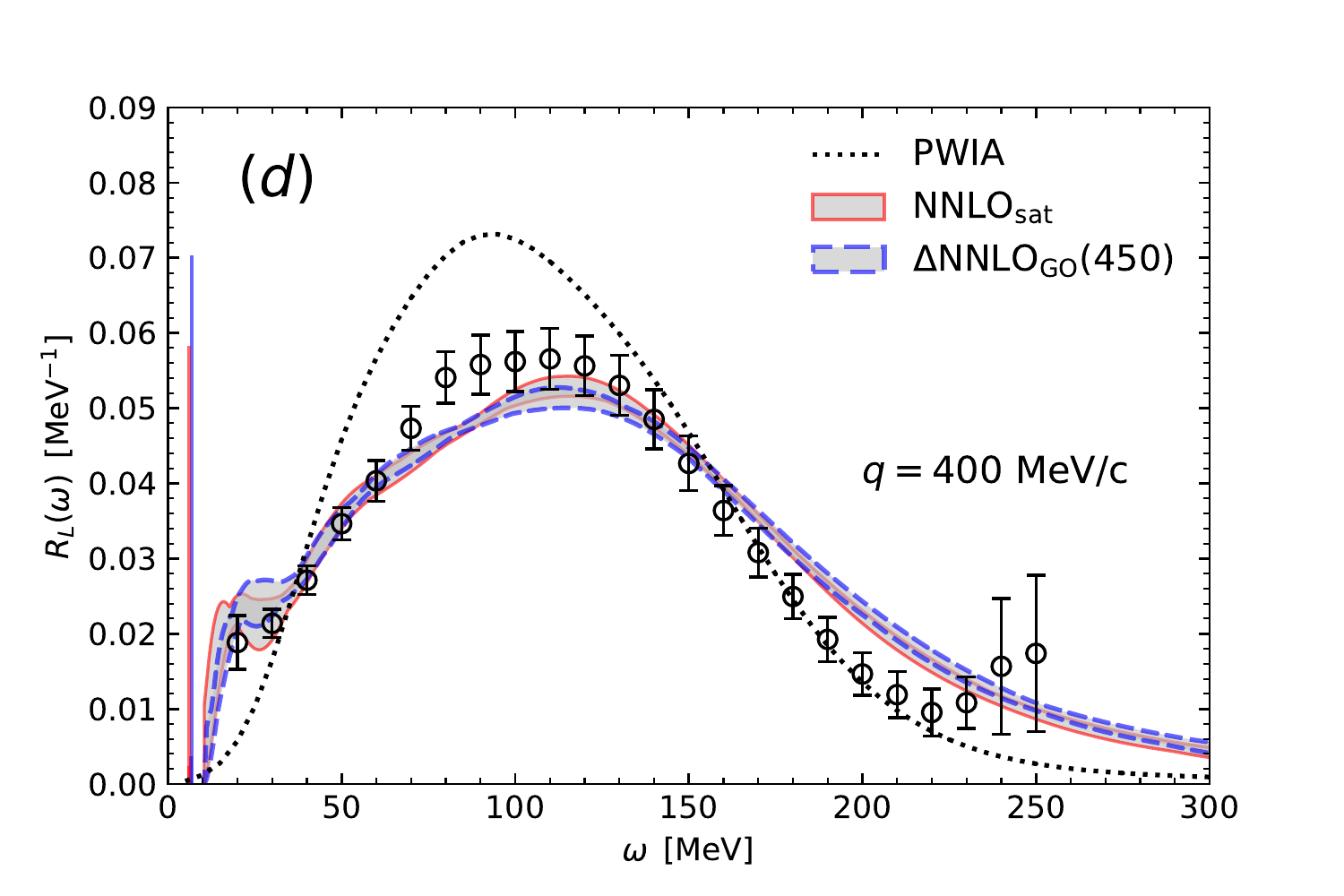}
	\caption{Longitudinal response of $^{40}$Ca for $q=300,\, 350,\, 400$ MeV/c for \NNLOsat\ and \NNLOgod\ potentials. For $q=200$ MeV/c the strength of excited states was quenched by factor of 2 for better visibility. Experimental data taken from~\cite{Williamson:1997zz}. }
	\label{40Ca}
\end{figure*}

{\it Benchmark on the $^{40}\mathrm{Ca}$ nucleus ---}
Following the same steps as in Ref.~\cite{Sobczyk:2020qtw},  we calculate the Coulomb sum rule for $^{40}$Ca using the \NNLOsat~ interaction.
We observe that the CoM contamination is negligible for $q>200$ MeV/c, and is overall much smaller than in the previously considered cases of $^4$He and $^{16}$O~\cite{Sobczyk:2020qtw}.
 In Fig.~\ref{CSR} we compare it to the cluster variational Monte Carlo (CVMC) results from Ref.~\cite{Lonardoni} which used the AV18+UIX potential and included Darwin-Foldy and spin-orbit corrections.
 Results are compatible at low-$q$ due to the larger uncertainty in the CVMC curve, and show the same increasing trend for $q>100$ MeV/c with small differences.  We have verified that the difference at $q=500$ MeV/c is mainly due to relativistic effects which we omitted in order to be consistent with the chiral order we work at. 
Most importantly, both theoretical predictions are in agreement with experimental data~\cite{Williamson:1997zz} in the range between 300 and 375 MeV/c and are higher than the data above $q=400$ MeV/c, likely because experimental data are obtained by integrating $R_L$ up to a finite $\omega$, and not up to infinity as is done in the theoretical calculations.
We consider this a successful benchmark of our method and point out that only a mild Hamiltonian dependence is observed.

{\it The $^{40}\mathrm{Ca}$ longitudinal response function ---}
We now turn to our \emph{ab initio} calculation of $R_L$ in $^{40}$Ca where the full final state interaction is considered. 
We choose $^{40}$Ca because we can compare our calculations with existing data, and it is also a stepping stone for coupled-cluster computations of neutrino scattering on $^{40}$Ar.
 For both \NNLOsat\ and \NNLOgod\ we perform computations of $R_L$ at the momentum transfers $q=200$, 300, 350 and 400 MeV/c.
 In CCSD, the obtained ground-state energies $E_0$ (proton separation energies $\omega_{th}$)  are $300.1$ (6.32) MeV and $322.12$ (6.12) MeV for the \NNLOsat\ and the \NNLOgod\ potential, respectively.
 
First, we find two bound excited $J^\pi=3^-$, $5^-$ states lying respectively at 4.5(3.8) MeV and 4.7(4.0) MeV with the NNLOsat(\NNLOgod) interactions, which are in reasonable agreement with experimental data at $3.7$~MeV $(J^\pi=3^-)$ and  at $4.5$ MeV $(J^\pi=5^-)$. We plot their strengths as a line
in Fig.~\ref{40Ca}, and we observe that  it decreases with $q$.
Second, for the continuum response we show a  band that reflects the uncertainty associated with the LIT inversion and  the model space, as we vary the harmonic oscillator frequency $\hbar\Omega $ from 18  to 20 and 22 MeV.
 As can be seen in Fig.~\ref{40Ca}, for each momentum transfer we observe a mild dependence on the interaction, the latter being stronger at $q=200$ MeV/c. Comparing to the available experimental data from Ref.~\cite{Williamson:1997zz}, we find a generally very good 
 agreement, which is best for $q=300$ MeV/c.
At $q=400$ MeV/c, we see a  quenching of the quasi-elastic peak and an enhancement in the tail with respect to experiment. We speculate that this could potentially be explained by relativistic boost effects~\cite{Rocco:2018tes} or by the fact that, especially at high $q$ and high $\omega$, we are reaching the limits of applicability of chiral effective field theory set by the regulator cutoff 450 MeV/c. 

Finally, to quantify the effect of the final state interaction, we will contrast the LIT-CC results with those of the simple plane wave impulse approximation (PWIA). The point-proton longitudinal response function is obtained in PWIA assuming one outgoing free proton with mass $m$ and a spectator (A-1)-system with mass $M_s$,
\begin{equation} 
  R_L^\mathrm{PWIA}(\omega,q)=\!\!\int\!\! d{\bf p}  \,n({\bf p})\, 
  \delta \left( \omega-\frac{(\! {\bf p} + {\bf q} )^2}{2
  m}-\frac{\bf p^2}{2 M_{s}} - \omega_{th}\! \right),
  \label{PWIA}
\end{equation}
and then augmented with nucleon electric form factors. Here $n({\bf p})$ represents the proton momentum distribution calculated from coupled-cluster theory using the \NNLOsat~interaction, where
 CoM corrections  are found to be  negligible~\cite{com_mom_distr}.
Unlike the LIT-CC results, the PWIA curves shown in  Fig.~\ref{40Ca} are in poor agreement with the data: $(i)$ they miss the quasi-elastic peak position by up to 20 MeV, $(ii)$ they overestimate considerably the quasi--elastic peak size
by up to 40$\%$ and $(iii)$
 and they do not fully account for the asymmetric shape of the response. The differences between the LIT-CC and the PWIA results are very strong at lower $\omega$, where we observe that even for the highest momentum transfers here considered $q=400$ MeV/c, we describe the experimental data very well. This highlights the importance of consistently including the final state interaction. 

In order to provide a  prediction  for future measurements as opposed to a sole postdiction of existing data, we have calculated also the $q=200$ MeV/c kinematics, where no data exist yet. While this low-$q$ range may be less important for neutrino physics, this is where we have the largest uncertainty band (range of low-$q$ and low-$\omega$). New precise data could provide important tests of the 
\emph{ab initio}
nuclear structure theory. 
 An experimental program in this direction is presently under development in Mainz~\cite{Doria}.

{\it Conclusions---} We performed an \emph{ab initio} calculation of the longitudinal response function of $^{40}$Ca and obtained very good agreement with existing data. Our results are a proof of principle that the LIT-CC method is suitable to deliver responses for lepton-nucleus scattering at the momentum transfers relevant for neutrino oscillation experiments. Consequently, we extended the reach of  consistent \emph{ab initio}  calculations of electromagnetic responses at intermediate momentum transfers into a region of medium-mass nuclei, which until now was limited to  systems with $A\le 12$. 

Our framework allows for quantification of uncertainties stemming from truncations of model space, chiral effective-field-theory, and coupled-cluster expansions. In this work, we estimated errors that arise from the inversion procedure, and studied the dependencies on the model space and the nuclear Hamiltonian.
Our quantified uncertainties does not yet include effects of missing higher-order excitations in the coupled-cluster expansion or terms in the chiral effective field theory interactions and currents.  A thorough analysis of all theory uncertainties  entering lepton-nucleus cross sections is part  of our future plans.

\begin{acknowledgments}
We thank Nir Barnea and Thomas Papenbrock for useful comments and discussions. This work was supported 
 by the Deutsche
Forschungsgemeinschaft (DFG) through the Collaborative Research Center
[The Low-Energy Frontier of the Standard Model (SFB 1044)], and
through the Cluster of Excellence ``Precision Physics, Fundamental
Interactions, and Structure of Matter" (PRISMA$^+$ EXC 2118/1) funded by the
DFG within the German Excellence Strategy (Project ID 39083149), by the
 Office of Nuclear Physics,
U.S. Department of Energy, under grants desc0018223 (NUCLEI SciDAC-4
collaboration) and by the Field Work Proposal ERKBP72 at Oak Ridge
National Laboratory (ORNL).
Computer time was provided by the
Innovative and Novel Computational Impact on Theory and Experiment
(INCITE) program and by the supercomputer Mogon at Johannes Gutenberg-Universit\"{a}t Mainz.
This research used resources of the Oak Ridge
Leadership Computing Facility located at ORNL, which is supported by
the Office of Science of the Department of Energy under Contract
No. DE-AC05-00OR22725.

\end{acknowledgments}

\FloatBarrier 
\bibliography{master,refs} 

\begin{thebibliography}{49}%
\makeatletter
\providecommand \@ifxundefined [1]{%
 \@ifx{#1\undefined}
}%
\providecommand \@ifnum [1]{%
 \ifnum #1\expandafter \@firstoftwo
 \else \expandafter \@secondoftwo
 \fi
}%
\providecommand \@ifx [1]{%
 \ifx #1\expandafter \@firstoftwo
 \else \expandafter \@secondoftwo
 \fi
}%
\providecommand \natexlab [1]{#1}%
\providecommand \enquote  [1]{``#1''}%
\providecommand \bibnamefont  [1]{#1}%
\providecommand \bibfnamefont [1]{#1}%
\providecommand \citenamefont [1]{#1}%
\providecommand \href@noop [0]{\@secondoftwo}%
\providecommand \href [0]{\begingroup \@sanitize@url \@href}%
\providecommand \@href[1]{\@@startlink{#1}\@@href}%
\providecommand \@@href[1]{\endgroup#1\@@endlink}%
\providecommand \@sanitize@url [0]{\catcode `\\12\catcode `\$12\catcode
  `\&12\catcode `\#12\catcode `\^12\catcode `\_12\catcode `\%12\relax}%
\providecommand \@@startlink[1]{}%
\providecommand \@@endlink[0]{}%
\providecommand \url  [0]{\begingroup\@sanitize@url \@url }%
\providecommand \@url [1]{\endgroup\@href {#1}{\urlprefix }}%
\providecommand \urlprefix  [0]{URL }%
\providecommand \Eprint [0]{\href }%
\providecommand \doibase [0]{http://dx.doi.org/}%
\providecommand \selectlanguage [0]{\@gobble}%
\providecommand \bibinfo  [0]{\@secondoftwo}%
\providecommand \bibfield  [0]{\@secondoftwo}%
\providecommand \translation [1]{[#1]}%
\providecommand \BibitemOpen [0]{}%
\providecommand \bibitemStop [0]{}%
\providecommand \bibitemNoStop [0]{.\EOS\space}%
\providecommand \EOS [0]{\spacefactor3000\relax}%
\providecommand \BibitemShut  [1]{\csname bibitem#1\endcsname}%
\let\auto@bib@innerbib\@empty
\bibitem [{\citenamefont {Hergert}(2020)}]{Heiko}%
  \BibitemOpen
  \bibfield  {author} {\bibinfo {author} {\bibfnamefont {Heiko}\ \bibnamefont
  {Hergert}},\ }\bibfield  {title} {\enquote {\bibinfo {title} {A guided tour
  of ab initio nuclear many-body theory},}\ }\href {\doibase
  10.3389/fphy.2020.00379} {\bibfield  {journal} {\bibinfo  {journal}
  {Frontiers in Physics}\ }\textbf {\bibinfo {volume} {8}},\ \bibinfo {pages}
  {379} (\bibinfo {year} {2020})},\ \bibinfo {note} {and references
  therein}\BibitemShut {NoStop}%
\bibitem [{\citenamefont {van Kolck}(1994)}]{vankolck1994}%
  \BibitemOpen
  \bibfield  {author} {\bibinfo {author} {\bibfnamefont {U.}~\bibnamefont {van
  Kolck}},\ }\bibfield  {title} {\enquote {\bibinfo {title} {{Few-nucleon
  forces from chiral Lagrangians}},}\ }\href {\doibase
  10.1103/PhysRevC.49.2932} {\bibfield  {journal} {\bibinfo  {journal} {Phys.
  Rev. C}\ }\textbf {\bibinfo {volume} {49}},\ \bibinfo {pages} {2932--2941}
  (\bibinfo {year} {1994})}\BibitemShut {NoStop}%
\bibitem [{\citenamefont {{Bedaque}}\ and\ \citenamefont {{van
  Kolck}}(2002)}]{bedaque2002}%
  \BibitemOpen
  \bibfield  {author} {\bibinfo {author} {\bibfnamefont {P.~F.}\ \bibnamefont
  {{Bedaque}}}\ and\ \bibinfo {author} {\bibfnamefont {U.}~\bibnamefont {{van
  Kolck}}},\ }\bibfield  {title} {\enquote {\bibinfo {title} {{Effective field
  theory for few-nucleon systems}},}\ }\href {\doibase
  10.1146/annurev.nucl.52.050102.090637} {\bibfield  {journal} {\bibinfo
  {journal} {Annual Review of Nuclear and Particle Science}\ }\textbf {\bibinfo
  {volume} {52}},\ \bibinfo {pages} {339--396} (\bibinfo {year} {2002})},\
  \Eprint {http://arxiv.org/abs/nucl-th/0203055} {nucl-th/0203055} \BibitemShut
  {NoStop}%
\bibitem [{\citenamefont {Epelbaum}\ \emph {et~al.}(2009)\citenamefont
  {Epelbaum}, \citenamefont {Hammer},\ and\ \citenamefont
  {Mei\ss{}ner}}]{epelbaum2009}%
  \BibitemOpen
  \bibfield  {author} {\bibinfo {author} {\bibfnamefont {E.}~\bibnamefont
  {Epelbaum}}, \bibinfo {author} {\bibfnamefont {H.-W.}\ \bibnamefont
  {Hammer}}, \ and\ \bibinfo {author} {\bibfnamefont {Ulf-G.}\ \bibnamefont
  {Mei\ss{}ner}},\ }\bibfield  {title} {\enquote {\bibinfo {title} {Modern
  theory of nuclear forces},}\ }\href {\doibase 10.1103/RevModPhys.81.1773}
  {\bibfield  {journal} {\bibinfo  {journal} {Rev. Mod. Phys.}\ }\textbf
  {\bibinfo {volume} {81}},\ \bibinfo {pages} {1773--1825} (\bibinfo {year}
  {2009})}\BibitemShut {NoStop}%
\bibitem [{\citenamefont {Machleidt}\ and\ \citenamefont
  {Entem}(2011)}]{machleidt2011}%
  \BibitemOpen
  \bibfield  {author} {\bibinfo {author} {\bibfnamefont {R.}~\bibnamefont
  {Machleidt}}\ and\ \bibinfo {author} {\bibfnamefont {D.R.}\ \bibnamefont
  {Entem}},\ }\bibfield  {title} {\enquote {\bibinfo {title} {Chiral effective
  field theory and nuclear forces},}\ }\href {\doibase
  10.1016/j.physrep.2011.02.001} {\bibfield  {journal} {\bibinfo  {journal}
  {Physics Reports}\ }\textbf {\bibinfo {volume} {503}},\ \bibinfo {pages} {1
  -- 75} (\bibinfo {year} {2011})}\BibitemShut {NoStop}%
\bibitem [{\citenamefont {Alvarez-Ruso}\ \emph {et~al.}(2018)\citenamefont
  {Alvarez-Ruso} \emph {et~al.}}]{Nustec}%
  \BibitemOpen
  \bibfield  {author} {\bibinfo {author} {\bibfnamefont {L.}~\bibnamefont
  {Alvarez-Ruso}} \emph {et~al.} (\bibinfo {collaboration} {NuSTEC}),\
  }\bibfield  {title} {\enquote {\bibinfo {title} {{NuSTEC White Paper: Status
  and challenges of neutrino\textendash{}nucleus scattering}},}\ }\href
  {\doibase 10.1016/j.ppnp.2018.01.006} {\bibfield  {journal} {\bibinfo
  {journal} {Prog. Part. Nucl. Phys.}\ }\textbf {\bibinfo {volume} {100}},\
  \bibinfo {pages} {1--68} (\bibinfo {year} {2018})},\ \Eprint
  {http://arxiv.org/abs/1706.03621} {arXiv:1706.03621 [hep-ph]} \BibitemShut
  {NoStop}%
\bibitem [{\citenamefont {Abe}\ \emph {et~al.}(2015)\citenamefont {Abe} \emph
  {et~al.}}]{hyperk}%
  \BibitemOpen
  \bibfield  {author} {\bibinfo {author} {\bibfnamefont {K.}~\bibnamefont
  {Abe}} \emph {et~al.} (\bibinfo {collaboration} {Hyper-Kamiokande
  Proto-Collaboration}),\ }\bibfield  {title} {\enquote {\bibinfo {title}
  {{Physics potential of a long-baseline neutrino oscillation experiment using
  a J-PARC neutrino beam and Hyper-Kamiokande}},}\ }\href {\doibase
  10.1093/ptep/ptv061} {\bibfield  {journal} {\bibinfo  {journal} {PTEP}\
  }\textbf {\bibinfo {volume} {2015}},\ \bibinfo {pages} {053C02} (\bibinfo
  {year} {2015})}\BibitemShut {NoStop}%
\bibitem [{\citenamefont {Acciarri}\ \emph {et~al.}(2015)\citenamefont
  {Acciarri} \emph {et~al.}}]{DUNE}%
  \BibitemOpen
  \bibfield  {author} {\bibinfo {author} {\bibfnamefont {R.}~\bibnamefont
  {Acciarri}} \emph {et~al.} (\bibinfo {collaboration} {DUNE}),\ }\bibfield
  {title} {\enquote {\bibinfo {title} {{Long-Baseline Neutrino Facility (LBNF)
  and Deep Underground Neutrino Experiment (DUNE)}},}\ }\href@noop {} {\
  (\bibinfo {year} {2015})},\ \Eprint {http://arxiv.org/abs/1512.06148}
  {arXiv:1512.06148 [physics.ins-det]} \BibitemShut {NoStop}%
\bibitem [{\citenamefont {Lovato}\ \emph {et~al.}(2014)\citenamefont {Lovato},
  \citenamefont {Gandolfi}, \citenamefont {Carlson}, \citenamefont {Pieper},\
  and\ \citenamefont {Schiavilla}}]{Lovato2014}%
  \BibitemOpen
  \bibfield  {author} {\bibinfo {author} {\bibfnamefont {A.}~\bibnamefont
  {Lovato}}, \bibinfo {author} {\bibfnamefont {S.}~\bibnamefont {Gandolfi}},
  \bibinfo {author} {\bibfnamefont {J.}~\bibnamefont {Carlson}}, \bibinfo
  {author} {\bibfnamefont {Steven~C.}\ \bibnamefont {Pieper}}, \ and\ \bibinfo
  {author} {\bibfnamefont {R.}~\bibnamefont {Schiavilla}},\ }\bibfield  {title}
  {\enquote {\bibinfo {title} {Neutral weak current two-body contributions in
  inclusive scattering from $^{12}\mathrm{C}$},}\ }\href {\doibase
  10.1103/PhysRevLett.112.182502} {\bibfield  {journal} {\bibinfo  {journal}
  {Phys. Rev. Lett.}\ }\textbf {\bibinfo {volume} {112}},\ \bibinfo {pages}
  {182502} (\bibinfo {year} {2014})}\BibitemShut {NoStop}%
\bibitem [{\citenamefont {Lovato}\ \emph {et~al.}(2015)\citenamefont {Lovato},
  \citenamefont {Gandolfi}, \citenamefont {Carlson}, \citenamefont {Pieper},\
  and\ \citenamefont {Schiavilla}}]{Lovato:2015qka}%
  \BibitemOpen
  \bibfield  {author} {\bibinfo {author} {\bibfnamefont {A.}~\bibnamefont
  {Lovato}}, \bibinfo {author} {\bibfnamefont {S.}~\bibnamefont {Gandolfi}},
  \bibinfo {author} {\bibfnamefont {J.}~\bibnamefont {Carlson}}, \bibinfo
  {author} {\bibfnamefont {Steven~C.}\ \bibnamefont {Pieper}}, \ and\ \bibinfo
  {author} {\bibfnamefont {R.}~\bibnamefont {Schiavilla}},\ }\bibfield  {title}
  {\enquote {\bibinfo {title} {{Electromagnetic and neutral-weak response
  functions of $^4$He and $^{12}$C}},}\ }\href {\doibase
  10.1103/PhysRevC.91.062501} {\bibfield  {journal} {\bibinfo  {journal} {Phys.
  Rev. C}\ }\textbf {\bibinfo {volume} {91}},\ \bibinfo {pages} {062501}
  (\bibinfo {year} {2015})},\ \Eprint {http://arxiv.org/abs/1501.01981}
  {arXiv:1501.01981 [nucl-th]} \BibitemShut {NoStop}%
\bibitem [{\citenamefont {Lovato}\ \emph {et~al.}(2018)\citenamefont {Lovato},
  \citenamefont {Gandolfi}, \citenamefont {Carlson}, \citenamefont {Lusk},
  \citenamefont {Pieper},\ and\ \citenamefont {Schiavilla}}]{Lovato:2017cux}%
  \BibitemOpen
  \bibfield  {author} {\bibinfo {author} {\bibfnamefont {A.}~\bibnamefont
  {Lovato}}, \bibinfo {author} {\bibfnamefont {S.}~\bibnamefont {Gandolfi}},
  \bibinfo {author} {\bibfnamefont {J.}~\bibnamefont {Carlson}}, \bibinfo
  {author} {\bibfnamefont {Ewing}\ \bibnamefont {Lusk}}, \bibinfo {author}
  {\bibfnamefont {Steven~C.}\ \bibnamefont {Pieper}}, \ and\ \bibinfo {author}
  {\bibfnamefont {R.}~\bibnamefont {Schiavilla}},\ }\bibfield  {title}
  {\enquote {\bibinfo {title} {{Quantum Monte Carlo calculation of
  neutral-current $\nu-^{12}C$ inclusive quasielastic scattering}},}\ }\href
  {\doibase 10.1103/PhysRevC.97.022502} {\bibfield  {journal} {\bibinfo
  {journal} {Phys. Rev. C}\ }\textbf {\bibinfo {volume} {97}},\ \bibinfo
  {pages} {022502} (\bibinfo {year} {2018})},\ \Eprint
  {http://arxiv.org/abs/1711.02047} {arXiv:1711.02047 [nucl-th]} \BibitemShut
  {NoStop}%
\bibitem [{\citenamefont {Lovato}\ \emph {et~al.}(2020)\citenamefont {Lovato},
  \citenamefont {Carlson}, \citenamefont {Gandolfi}, \citenamefont {Rocco},\
  and\ \citenamefont {Schiavilla}}]{lovato2020}%
  \BibitemOpen
  \bibfield  {author} {\bibinfo {author} {\bibfnamefont {A.}~\bibnamefont
  {Lovato}}, \bibinfo {author} {\bibfnamefont {J.}~\bibnamefont {Carlson}},
  \bibinfo {author} {\bibfnamefont {S.}~\bibnamefont {Gandolfi}}, \bibinfo
  {author} {\bibfnamefont {N.}~\bibnamefont {Rocco}}, \ and\ \bibinfo {author}
  {\bibfnamefont {R.}~\bibnamefont {Schiavilla}},\ }\bibfield  {title}
  {\enquote {\bibinfo {title} {Ab initio study of
  $({\ensuremath{\nu}}_{\ensuremath{\ell}},{\ensuremath{\ell}}^{\ensuremath{-}})$
  and
  $({\overline{\ensuremath{\nu}}}_{\ensuremath{\ell}},{\ensuremath{\ell}}^{+})$
  inclusive scattering in $^{12}\mathrm{C}$: Confronting the miniboone and t2k
  ccqe data},}\ }\href {\doibase 10.1103/PhysRevX.10.031068} {\bibfield
  {journal} {\bibinfo  {journal} {Phys. Rev. X}\ }\textbf {\bibinfo {volume}
  {10}},\ \bibinfo {pages} {031068} (\bibinfo {year} {2020})}\BibitemShut
  {NoStop}%
\bibitem [{\citenamefont {Pastore}\ \emph {et~al.}(2020)\citenamefont
  {Pastore}, \citenamefont {Carlson}, \citenamefont {Gandolfi}, \citenamefont
  {Schiavilla},\ and\ \citenamefont {Wiringa}}]{STA}%
  \BibitemOpen
  \bibfield  {author} {\bibinfo {author} {\bibfnamefont {S.}~\bibnamefont
  {Pastore}}, \bibinfo {author} {\bibfnamefont {J.}~\bibnamefont {Carlson}},
  \bibinfo {author} {\bibfnamefont {S.}~\bibnamefont {Gandolfi}}, \bibinfo
  {author} {\bibfnamefont {R.}~\bibnamefont {Schiavilla}}, \ and\ \bibinfo
  {author} {\bibfnamefont {R.~B.}\ \bibnamefont {Wiringa}},\ }\bibfield
  {title} {\enquote {\bibinfo {title} {Quasielastic lepton scattering and
  back-to-back nucleons in the short-time approximation},}\ }\href {\doibase
  10.1103/PhysRevC.101.044612} {\bibfield  {journal} {\bibinfo  {journal}
  {Phys. Rev. C}\ }\textbf {\bibinfo {volume} {101}},\ \bibinfo {pages}
  {044612} (\bibinfo {year} {2020})}\BibitemShut {NoStop}%
\bibitem [{\citenamefont {Rocco}\ and\ \citenamefont
  {Barbieri}(2018)}]{Rocco:2018vbf}%
  \BibitemOpen
  \bibfield  {author} {\bibinfo {author} {\bibfnamefont {N.}~\bibnamefont
  {Rocco}}\ and\ \bibinfo {author} {\bibfnamefont {C.}~\bibnamefont
  {Barbieri}},\ }\bibfield  {title} {\enquote {\bibinfo {title} {{Inclusive
  electron-nucleus cross section within the Self Consistent Green's Function
  approach}},}\ }\href {\doibase 10.1103/PhysRevC.98.025501} {\bibfield
  {journal} {\bibinfo  {journal} {Phys. Rev. C}\ }\textbf {\bibinfo {volume}
  {98}},\ \bibinfo {pages} {025501} (\bibinfo {year} {2018})},\ \Eprint
  {http://arxiv.org/abs/1803.00825} {arXiv:1803.00825 [nucl-th]} \BibitemShut
  {NoStop}%
\bibitem [{\citenamefont {Barbieri}\ \emph {et~al.}(2019)\citenamefont
  {Barbieri}, \citenamefont {Rocco},\ and\ \citenamefont
  {Som\`{a}}}]{Barbieri:2019ual}%
  \BibitemOpen
  \bibfield  {author} {\bibinfo {author} {\bibfnamefont {C.}~\bibnamefont
  {Barbieri}}, \bibinfo {author} {\bibfnamefont {N.}~\bibnamefont {Rocco}}, \
  and\ \bibinfo {author} {\bibfnamefont {V.}~\bibnamefont {Som\`{a}}},\
  }\bibfield  {title} {\enquote {\bibinfo {title} {{Lepton scattering from
  $^{40}$Ar and Ti in the quasielastic peak region}},}\ }\href@noop {} {\
  (\bibinfo {year} {2019})},\ \Eprint {http://arxiv.org/abs/1907.01122}
  {arXiv:1907.01122 [nucl-th]} \BibitemShut {NoStop}%
\bibitem [{\citenamefont {Coester}(1958)}]{coester1958}%
  \BibitemOpen
  \bibfield  {author} {\bibinfo {author} {\bibfnamefont {F.}~\bibnamefont
  {Coester}},\ }\bibfield  {title} {\enquote {\bibinfo {title} {Bound states of
  a many-particle system},}\ }\href {\doibase 10.1016/0029-5582(58)90280-3}
  {\bibfield  {journal} {\bibinfo  {journal} {Nuclear Physics}\ }\textbf
  {\bibinfo {volume} {7}},\ \bibinfo {pages} {421 -- 424} (\bibinfo {year}
  {1958})}\BibitemShut {NoStop}%
\bibitem [{\citenamefont {Coester}\ and\ \citenamefont
  {K{\"u}mmel}(1960)}]{coester1960}%
  \BibitemOpen
  \bibfield  {author} {\bibinfo {author} {\bibfnamefont {F.}~\bibnamefont
  {Coester}}\ and\ \bibinfo {author} {\bibfnamefont {H.}~\bibnamefont
  {K{\"u}mmel}},\ }\bibfield  {title} {\enquote {\bibinfo {title} {Short-range
  correlations in nuclear wave functions},}\ }\href {\doibase
  10.1016/0029-5582(60)90140-1} {\bibfield  {journal} {\bibinfo  {journal}
  {Nuclear Physics}\ }\textbf {\bibinfo {volume} {17}},\ \bibinfo {pages} {477
  -- 485} (\bibinfo {year} {1960})}\BibitemShut {NoStop}%
\bibitem [{\citenamefont {K{\"u}mmel}\ \emph {et~al.}(1978)\citenamefont
  {K{\"u}mmel}, \citenamefont {L{\"u}hrmann},\ and\ \citenamefont
  {Zabolitzky}}]{kuemmel1978}%
  \BibitemOpen
  \bibfield  {author} {\bibinfo {author} {\bibfnamefont {H.}~\bibnamefont
  {K{\"u}mmel}}, \bibinfo {author} {\bibfnamefont {K.~H.}\ \bibnamefont
  {L{\"u}hrmann}}, \ and\ \bibinfo {author} {\bibfnamefont {J.~G.}\
  \bibnamefont {Zabolitzky}},\ }\bibfield  {title} {\enquote {\bibinfo {title}
  {{Many-fermion theory in expS- (or coupled cluster) form}},}\ }\href
  {\doibase 10.1016/0370-1573(78)90081-9} {\bibfield  {journal} {\bibinfo
  {journal} {Physics Reports}\ }\textbf {\bibinfo {volume} {36}},\ \bibinfo
  {pages} {1 -- 63} (\bibinfo {year} {1978})}\BibitemShut {NoStop}%
\bibitem [{\citenamefont {Mihaila}\ and\ \citenamefont
  {Heisenberg}(2000)}]{mihaila2000b}%
  \BibitemOpen
  \bibfield  {author} {\bibinfo {author} {\bibfnamefont {B.}~\bibnamefont
  {Mihaila}}\ and\ \bibinfo {author} {\bibfnamefont {J.~H.}\ \bibnamefont
  {Heisenberg}},\ }\bibfield  {title} {\enquote {\bibinfo {title} {{Microscopic
  Calculation of the Inclusive Electron Scattering Structure Function in
  $^{16}O$}},}\ }\href {\doibase 10.1103/PhysRevLett.84.1403} {\bibfield
  {journal} {\bibinfo  {journal} {Phys. Rev. Lett.}\ }\textbf {\bibinfo
  {volume} {84}},\ \bibinfo {pages} {1403--1406} (\bibinfo {year}
  {2000})}\BibitemShut {NoStop}%
\bibitem [{\citenamefont {Dean}\ and\ \citenamefont
  {Hjorth-Jensen}(2004)}]{dean2004}%
  \BibitemOpen
  \bibfield  {author} {\bibinfo {author} {\bibfnamefont {D.~J.}\ \bibnamefont
  {Dean}}\ and\ \bibinfo {author} {\bibfnamefont {M.}~\bibnamefont
  {Hjorth-Jensen}},\ }\bibfield  {title} {\enquote {\bibinfo {title}
  {Coupled-cluster approach to nuclear physics},}\ }\href {\doibase
  10.1103/PhysRevC.69.054320} {\bibfield  {journal} {\bibinfo  {journal} {Phys.
  Rev. C}\ }\textbf {\bibinfo {volume} {69}},\ \bibinfo {pages} {054320}
  (\bibinfo {year} {2004})}\BibitemShut {NoStop}%
\bibitem [{\citenamefont {W\l{}och}\ \emph {et~al.}(2005)\citenamefont
  {W\l{}och}, \citenamefont {Dean}, \citenamefont {Gour}, \citenamefont
  {Hjorth-Jensen}, \citenamefont {Kowalski}, \citenamefont {Papenbrock},\ and\
  \citenamefont {Piecuch}}]{wloch2005}%
  \BibitemOpen
  \bibfield  {author} {\bibinfo {author} {\bibfnamefont {M.}~\bibnamefont
  {W\l{}och}}, \bibinfo {author} {\bibfnamefont {D.~J.}\ \bibnamefont {Dean}},
  \bibinfo {author} {\bibfnamefont {J.~R.}\ \bibnamefont {Gour}}, \bibinfo
  {author} {\bibfnamefont {M.}~\bibnamefont {Hjorth-Jensen}}, \bibinfo {author}
  {\bibfnamefont {K.}~\bibnamefont {Kowalski}}, \bibinfo {author}
  {\bibfnamefont {T.}~\bibnamefont {Papenbrock}}, \ and\ \bibinfo {author}
  {\bibfnamefont {P.}~\bibnamefont {Piecuch}},\ }\bibfield  {title} {\enquote
  {\bibinfo {title} {\textit{Ab-Initio} coupled-cluster study of
  $^{16}\mathrm{O}$},}\ }\href {\doibase 10.1103/PhysRevLett.94.212501}
  {\bibfield  {journal} {\bibinfo  {journal} {Phys. Rev. Lett.}\ }\textbf
  {\bibinfo {volume} {94}},\ \bibinfo {pages} {212501} (\bibinfo {year}
  {2005})}\BibitemShut {NoStop}%
\bibitem [{\citenamefont {Hagen}\ \emph {et~al.}(2008)\citenamefont {Hagen},
  \citenamefont {Papenbrock}, \citenamefont {Dean},\ and\ \citenamefont
  {Hjorth-Jensen}}]{hagen2008}%
  \BibitemOpen
  \bibfield  {author} {\bibinfo {author} {\bibfnamefont {G.}~\bibnamefont
  {Hagen}}, \bibinfo {author} {\bibfnamefont {T.}~\bibnamefont {Papenbrock}},
  \bibinfo {author} {\bibfnamefont {D.~J.}\ \bibnamefont {Dean}}, \ and\
  \bibinfo {author} {\bibfnamefont {M.}~\bibnamefont {Hjorth-Jensen}},\
  }\bibfield  {title} {\enquote {\bibinfo {title} {Medium-mass nuclei from
  chiral nucleon-nucleon interactions},}\ }\href {\doibase
  10.1103/PhysRevLett.101.092502} {\bibfield  {journal} {\bibinfo  {journal}
  {Phys. Rev. Lett.}\ }\textbf {\bibinfo {volume} {101}},\ \bibinfo {pages}
  {092502} (\bibinfo {year} {2008})}\BibitemShut {NoStop}%
\bibitem [{\citenamefont {Hagen}\ \emph {et~al.}(2010)\citenamefont {Hagen},
  \citenamefont {Papenbrock}, \citenamefont {Dean},\ and\ \citenamefont
  {Hjorth-Jensen}}]{hagen2010b}%
  \BibitemOpen
  \bibfield  {author} {\bibinfo {author} {\bibfnamefont {G.}~\bibnamefont
  {Hagen}}, \bibinfo {author} {\bibfnamefont {T.}~\bibnamefont {Papenbrock}},
  \bibinfo {author} {\bibfnamefont {D.~J.}\ \bibnamefont {Dean}}, \ and\
  \bibinfo {author} {\bibfnamefont {M.}~\bibnamefont {Hjorth-Jensen}},\
  }\bibfield  {title} {\enquote {\bibinfo {title} {\textit{Ab initio}
  coupled-cluster approach to nuclear structure with modern nucleon-nucleon
  interactions},}\ }\href {\doibase 10.1103/PhysRevC.82.034330} {\bibfield
  {journal} {\bibinfo  {journal} {Phys. Rev. C}\ }\textbf {\bibinfo {volume}
  {82}},\ \bibinfo {pages} {034330} (\bibinfo {year} {2010})}\BibitemShut
  {NoStop}%
\bibitem [{\citenamefont {Binder}\ \emph {et~al.}(2014)\citenamefont {Binder},
  \citenamefont {Langhammer}, \citenamefont {Calci},\ and\ \citenamefont
  {Roth}}]{binder2013b}%
  \BibitemOpen
  \bibfield  {author} {\bibinfo {author} {\bibfnamefont {Sven}\ \bibnamefont
  {Binder}}, \bibinfo {author} {\bibfnamefont {Joachim}\ \bibnamefont
  {Langhammer}}, \bibinfo {author} {\bibfnamefont {Angelo}\ \bibnamefont
  {Calci}}, \ and\ \bibinfo {author} {\bibfnamefont {Robert}\ \bibnamefont
  {Roth}},\ }\bibfield  {title} {\enquote {\bibinfo {title} {Ab initio path to
  heavy nuclei},}\ }\href {\doibase 10.1016/j.physletb.2014.07.010} {\bibfield
  {journal} {\bibinfo  {journal} {Phys. Lett. B}\ }\textbf {\bibinfo {volume}
  {736}},\ \bibinfo {pages} {119 -- 123} (\bibinfo {year} {2014})}\BibitemShut
  {NoStop}%
\bibitem [{\citenamefont {Hagen}\ \emph {et~al.}(2014)\citenamefont {Hagen},
  \citenamefont {Papenbrock}, \citenamefont {Hjorth-Jensen},\ and\
  \citenamefont {Dean}}]{hagen2014}%
  \BibitemOpen
  \bibfield  {author} {\bibinfo {author} {\bibfnamefont {G.}~\bibnamefont
  {Hagen}}, \bibinfo {author} {\bibfnamefont {T.}~\bibnamefont {Papenbrock}},
  \bibinfo {author} {\bibfnamefont {M.}~\bibnamefont {Hjorth-Jensen}}, \ and\
  \bibinfo {author} {\bibfnamefont {D.~J.}\ \bibnamefont {Dean}},\ }\bibfield
  {title} {\enquote {\bibinfo {title} {Coupled-cluster computations of atomic
  nuclei},}\ }\href {\doibase 10.1088/0034-4885/77/9/096302} {\bibfield
  {journal} {\bibinfo  {journal} {Rep. Prog. Phys.}\ }\textbf {\bibinfo
  {volume} {77}},\ \bibinfo {pages} {096302} (\bibinfo {year}
  {2014})}\BibitemShut {NoStop}%
\bibitem [{\citenamefont {Liu}\ \emph {et~al.}(2019)\citenamefont {Liu},
  \citenamefont {Obertelli}, \citenamefont {Doornenbal}, \citenamefont
  {Bertulani}, \citenamefont {Hagen}, \citenamefont {Holt}, \citenamefont
  {Jansen}, \citenamefont {Morris}, \citenamefont {Schwenk}, \citenamefont
  {Stroberg}, \citenamefont {Achouri}, \citenamefont {Baba}, \citenamefont
  {Browne}, \citenamefont {Calvet}, \citenamefont {Ch\^ateau}, \citenamefont
  {Chen}, \citenamefont {Chiga}, \citenamefont {Corsi}, \citenamefont
  {Cort\'es}, \citenamefont {Delbart}, \citenamefont {Gheller}, \citenamefont
  {Giganon}, \citenamefont {Gillibert}, \citenamefont {Hilaire}, \citenamefont
  {Isobe}, \citenamefont {Kobayashi}, \citenamefont {Kubota}, \citenamefont
  {Lapoux}, \citenamefont {Motobayashi}, \citenamefont {Murray}, \citenamefont
  {Otsu}, \citenamefont {Panin}, \citenamefont {Paul}, \citenamefont
  {Rodriguez}, \citenamefont {Sakurai}, \citenamefont {Sasano}, \citenamefont
  {Steppenbeck}, \citenamefont {Stuhl}, \citenamefont {Sun}, \citenamefont
  {Togano}, \citenamefont {Uesaka}, \citenamefont {Wimmer}, \citenamefont
  {Yoneda}, \citenamefont {Aktas}, \citenamefont {Aumann}, \citenamefont
  {Chung}, \citenamefont {Flavigny}, \citenamefont {Franchoo}, \citenamefont
  {Ga\ifmmode \check{s}\else \v{s}\fi{}pari\ifmmode~\acute{c}\else \'{c}\fi{}},
  \citenamefont {Gerst}, \citenamefont {Gibelin}, \citenamefont {Hahn},
  \citenamefont {Kim}, \citenamefont {Koiwai}, \citenamefont {Kondo},
  \citenamefont {Koseoglou}, \citenamefont {Lee}, \citenamefont {Lehr},
  \citenamefont {Linh}, \citenamefont {Lokotko}, \citenamefont {MacCormick},
  \citenamefont {Moschner}, \citenamefont {Nakamura}, \citenamefont {Park},
  \citenamefont {Rossi}, \citenamefont {Sahin}, \citenamefont {Sohler},
  \citenamefont {S\"oderstr\"om}, \citenamefont {Takeuchi}, \citenamefont
  {T\"ornqvist}, \citenamefont {Vaquero}, \citenamefont {Wagner}, \citenamefont
  {Wang}, \citenamefont {Werner}, \citenamefont {Xu}, \citenamefont {Yamada},
  \citenamefont {Yan}, \citenamefont {Yang}, \citenamefont {Yasuda},\ and\
  \citenamefont {Zanetti}}]{liu2019}%
  \BibitemOpen
  \bibfield  {author} {\bibinfo {author} {\bibfnamefont {H.~N.}\ \bibnamefont
  {Liu}}, \bibinfo {author} {\bibfnamefont {A.}~\bibnamefont {Obertelli}},
  \bibinfo {author} {\bibfnamefont {P.}~\bibnamefont {Doornenbal}}, \bibinfo
  {author} {\bibfnamefont {C.~A.}\ \bibnamefont {Bertulani}}, \bibinfo {author}
  {\bibfnamefont {G.}~\bibnamefont {Hagen}}, \bibinfo {author} {\bibfnamefont
  {J.~D.}\ \bibnamefont {Holt}}, \bibinfo {author} {\bibfnamefont {G.~R.}\
  \bibnamefont {Jansen}}, \bibinfo {author} {\bibfnamefont {T.~D.}\
  \bibnamefont {Morris}}, \bibinfo {author} {\bibfnamefont {A.}~\bibnamefont
  {Schwenk}}, \bibinfo {author} {\bibfnamefont {R.}~\bibnamefont {Stroberg}},
  \bibinfo {author} {\bibfnamefont {N.}~\bibnamefont {Achouri}}, \bibinfo
  {author} {\bibfnamefont {H.}~\bibnamefont {Baba}}, \bibinfo {author}
  {\bibfnamefont {F.}~\bibnamefont {Browne}}, \bibinfo {author} {\bibfnamefont
  {D.}~\bibnamefont {Calvet}}, \bibinfo {author} {\bibfnamefont
  {F.}~\bibnamefont {Ch\^ateau}}, \bibinfo {author} {\bibfnamefont
  {S.}~\bibnamefont {Chen}}, \bibinfo {author} {\bibfnamefont {N.}~\bibnamefont
  {Chiga}}, \bibinfo {author} {\bibfnamefont {A.}~\bibnamefont {Corsi}},
  \bibinfo {author} {\bibfnamefont {M.~L.}\ \bibnamefont {Cort\'es}}, \bibinfo
  {author} {\bibfnamefont {A.}~\bibnamefont {Delbart}}, \bibinfo {author}
  {\bibfnamefont {J.-M.}\ \bibnamefont {Gheller}}, \bibinfo {author}
  {\bibfnamefont {A.}~\bibnamefont {Giganon}}, \bibinfo {author} {\bibfnamefont
  {A.}~\bibnamefont {Gillibert}}, \bibinfo {author} {\bibfnamefont
  {C.}~\bibnamefont {Hilaire}}, \bibinfo {author} {\bibfnamefont
  {T.}~\bibnamefont {Isobe}}, \bibinfo {author} {\bibfnamefont
  {T.}~\bibnamefont {Kobayashi}}, \bibinfo {author} {\bibfnamefont
  {Y.}~\bibnamefont {Kubota}}, \bibinfo {author} {\bibfnamefont
  {V.}~\bibnamefont {Lapoux}}, \bibinfo {author} {\bibfnamefont
  {T.}~\bibnamefont {Motobayashi}}, \bibinfo {author} {\bibfnamefont
  {I.}~\bibnamefont {Murray}}, \bibinfo {author} {\bibfnamefont
  {H.}~\bibnamefont {Otsu}}, \bibinfo {author} {\bibfnamefont {V.}~\bibnamefont
  {Panin}}, \bibinfo {author} {\bibfnamefont {N.}~\bibnamefont {Paul}},
  \bibinfo {author} {\bibfnamefont {W.}~\bibnamefont {Rodriguez}}, \bibinfo
  {author} {\bibfnamefont {H.}~\bibnamefont {Sakurai}}, \bibinfo {author}
  {\bibfnamefont {M.}~\bibnamefont {Sasano}}, \bibinfo {author} {\bibfnamefont
  {D.}~\bibnamefont {Steppenbeck}}, \bibinfo {author} {\bibfnamefont
  {L.}~\bibnamefont {Stuhl}}, \bibinfo {author} {\bibfnamefont {Y.~L.}\
  \bibnamefont {Sun}}, \bibinfo {author} {\bibfnamefont {Y.}~\bibnamefont
  {Togano}}, \bibinfo {author} {\bibfnamefont {T.}~\bibnamefont {Uesaka}},
  \bibinfo {author} {\bibfnamefont {K.}~\bibnamefont {Wimmer}}, \bibinfo
  {author} {\bibfnamefont {K.}~\bibnamefont {Yoneda}}, \bibinfo {author}
  {\bibfnamefont {O.}~\bibnamefont {Aktas}}, \bibinfo {author} {\bibfnamefont
  {T.}~\bibnamefont {Aumann}}, \bibinfo {author} {\bibfnamefont {L.~X.}\
  \bibnamefont {Chung}}, \bibinfo {author} {\bibfnamefont {F.}~\bibnamefont
  {Flavigny}}, \bibinfo {author} {\bibfnamefont {S.}~\bibnamefont {Franchoo}},
  \bibinfo {author} {\bibfnamefont {I.}~\bibnamefont {Ga\ifmmode \check{s}\else
  \v{s}\fi{}pari\ifmmode~\acute{c}\else \'{c}\fi{}}}, \bibinfo {author}
  {\bibfnamefont {R.-B.}\ \bibnamefont {Gerst}}, \bibinfo {author}
  {\bibfnamefont {J.}~\bibnamefont {Gibelin}}, \bibinfo {author} {\bibfnamefont
  {K.~I.}\ \bibnamefont {Hahn}}, \bibinfo {author} {\bibfnamefont
  {D.}~\bibnamefont {Kim}}, \bibinfo {author} {\bibfnamefont {T.}~\bibnamefont
  {Koiwai}}, \bibinfo {author} {\bibfnamefont {Y.}~\bibnamefont {Kondo}},
  \bibinfo {author} {\bibfnamefont {P.}~\bibnamefont {Koseoglou}}, \bibinfo
  {author} {\bibfnamefont {J.}~\bibnamefont {Lee}}, \bibinfo {author}
  {\bibfnamefont {C.}~\bibnamefont {Lehr}}, \bibinfo {author} {\bibfnamefont
  {B.~D.}\ \bibnamefont {Linh}}, \bibinfo {author} {\bibfnamefont
  {T.}~\bibnamefont {Lokotko}}, \bibinfo {author} {\bibfnamefont
  {M.}~\bibnamefont {MacCormick}}, \bibinfo {author} {\bibfnamefont
  {K.}~\bibnamefont {Moschner}}, \bibinfo {author} {\bibfnamefont
  {T.}~\bibnamefont {Nakamura}}, \bibinfo {author} {\bibfnamefont {S.~Y.}\
  \bibnamefont {Park}}, \bibinfo {author} {\bibfnamefont {D.}~\bibnamefont
  {Rossi}}, \bibinfo {author} {\bibfnamefont {E.}~\bibnamefont {Sahin}},
  \bibinfo {author} {\bibfnamefont {D.}~\bibnamefont {Sohler}}, \bibinfo
  {author} {\bibfnamefont {P.-A.}\ \bibnamefont {S\"oderstr\"om}}, \bibinfo
  {author} {\bibfnamefont {S.}~\bibnamefont {Takeuchi}}, \bibinfo {author}
  {\bibfnamefont {H.}~\bibnamefont {T\"ornqvist}}, \bibinfo {author}
  {\bibfnamefont {V.}~\bibnamefont {Vaquero}}, \bibinfo {author} {\bibfnamefont
  {V.}~\bibnamefont {Wagner}}, \bibinfo {author} {\bibfnamefont
  {S.}~\bibnamefont {Wang}}, \bibinfo {author} {\bibfnamefont {V.}~\bibnamefont
  {Werner}}, \bibinfo {author} {\bibfnamefont {X.}~\bibnamefont {Xu}}, \bibinfo
  {author} {\bibfnamefont {H.}~\bibnamefont {Yamada}}, \bibinfo {author}
  {\bibfnamefont {D.}~\bibnamefont {Yan}}, \bibinfo {author} {\bibfnamefont
  {Z.}~\bibnamefont {Yang}}, \bibinfo {author} {\bibfnamefont {M.}~\bibnamefont
  {Yasuda}}, \ and\ \bibinfo {author} {\bibfnamefont {L.}~\bibnamefont
  {Zanetti}},\ }\bibfield  {title} {\enquote {\bibinfo {title} {How robust is
  the $n=34$ subshell closure? first spectroscopy of $^{52}\mathrm{Ar}$},}\
  }\href {\doibase 10.1103/PhysRevLett.122.072502} {\bibfield  {journal}
  {\bibinfo  {journal} {Phys. Rev. Lett.}\ }\textbf {\bibinfo {volume} {122}},\
  \bibinfo {pages} {072502} (\bibinfo {year} {2019})}\BibitemShut {NoStop}%
\bibitem [{\citenamefont {Payne}\ \emph {et~al.}(2019)\citenamefont {Payne},
  \citenamefont {Bacca}, \citenamefont {Hagen}, \citenamefont {Jiang},\ and\
  \citenamefont {Papenbrock}}]{Payne:2019wvy}%
  \BibitemOpen
  \bibfield  {author} {\bibinfo {author} {\bibfnamefont {C.~G.}\ \bibnamefont
  {Payne}}, \bibinfo {author} {\bibfnamefont {S.}~\bibnamefont {Bacca}},
  \bibinfo {author} {\bibfnamefont {G.}~\bibnamefont {Hagen}}, \bibinfo
  {author} {\bibfnamefont {W.}~\bibnamefont {Jiang}}, \ and\ \bibinfo {author}
  {\bibfnamefont {T.}~\bibnamefont {Papenbrock}},\ }\bibfield  {title}
  {\enquote {\bibinfo {title} {{Coherent elastic neutrino-nucleus scattering on
  $^{40}$Ar from first principles}},}\ }\href {\doibase
  10.1103/PhysRevC.100.061304} {\bibfield  {journal} {\bibinfo  {journal}
  {Phys. Rev. C}\ }\textbf {\bibinfo {volume} {100}},\ \bibinfo {pages}
  {061304} (\bibinfo {year} {2019})},\ \Eprint
  {http://arxiv.org/abs/1908.09739} {arXiv:1908.09739 [nucl-th]} \BibitemShut
  {NoStop}%
\bibitem [{\citenamefont {Novario}\ \emph {et~al.}(2020)\citenamefont
  {Novario}, \citenamefont {Hagen}, \citenamefont {Jansen},\ and\ \citenamefont
  {Papenbrock}}]{novario2020}%
  \BibitemOpen
  \bibfield  {author} {\bibinfo {author} {\bibfnamefont {S.~J.}\ \bibnamefont
  {Novario}}, \bibinfo {author} {\bibfnamefont {G.}~\bibnamefont {Hagen}},
  \bibinfo {author} {\bibfnamefont {G.~R.}\ \bibnamefont {Jansen}}, \ and\
  \bibinfo {author} {\bibfnamefont {T.}~\bibnamefont {Papenbrock}},\ }\bibfield
   {title} {\enquote {\bibinfo {title} {Charge radii of exotic neon and
  magnesium isotopes},}\ }\href {\doibase 10.1103/PhysRevC.102.051303}
  {\bibfield  {journal} {\bibinfo  {journal} {Phys. Rev. C}\ }\textbf {\bibinfo
  {volume} {102}},\ \bibinfo {pages} {051303} (\bibinfo {year}
  {2020})}\BibitemShut {NoStop}%
\bibitem [{\citenamefont {Koszor{\'u}s}\ \emph {et~al.}(2021)\citenamefont
  {Koszor{\'u}s}, \citenamefont {Yang}, \citenamefont {Jiang}, \citenamefont
  {Novario}, \citenamefont {Bai}, \citenamefont {Billowes}, \citenamefont
  {Binnersley}, \citenamefont {Bissell}, \citenamefont {Cocolios},
  \citenamefont {Cooper}, \citenamefont {de~Groote}, \citenamefont
  {Ekstr{\"o}m}, \citenamefont {Flanagan}, \citenamefont {Forss{\'e}n},
  \citenamefont {Franchoo}, \citenamefont {Ruiz}, \citenamefont {Gustafsson},
  \citenamefont {Hagen}, \citenamefont {Jansen}, \citenamefont
  {Kanellakopoulos}, \citenamefont {Kortelainen}, \citenamefont {Nazarewicz},
  \citenamefont {Neyens}, \citenamefont {Papenbrock}, \citenamefont {Reinhard},
  \citenamefont {Ricketts}, \citenamefont {Sahoo}, \citenamefont {Vernon},\
  and\ \citenamefont {Wilkins}}]{koszorus2021}%
  \BibitemOpen
  \bibfield  {author} {\bibinfo {author} {\bibfnamefont {{\'A}.}~\bibnamefont
  {Koszor{\'u}s}}, \bibinfo {author} {\bibfnamefont {X.~F.}\ \bibnamefont
  {Yang}}, \bibinfo {author} {\bibfnamefont {W.~G.}\ \bibnamefont {Jiang}},
  \bibinfo {author} {\bibfnamefont {S.~J.}\ \bibnamefont {Novario}}, \bibinfo
  {author} {\bibfnamefont {S.~W.}\ \bibnamefont {Bai}}, \bibinfo {author}
  {\bibfnamefont {J.}~\bibnamefont {Billowes}}, \bibinfo {author}
  {\bibfnamefont {C.~L.}\ \bibnamefont {Binnersley}}, \bibinfo {author}
  {\bibfnamefont {M.~L.}\ \bibnamefont {Bissell}}, \bibinfo {author}
  {\bibfnamefont {T.~E.}\ \bibnamefont {Cocolios}}, \bibinfo {author}
  {\bibfnamefont {B.~S.}\ \bibnamefont {Cooper}}, \bibinfo {author}
  {\bibfnamefont {R.~P.}\ \bibnamefont {de~Groote}}, \bibinfo {author}
  {\bibfnamefont {A.}~\bibnamefont {Ekstr{\"o}m}}, \bibinfo {author}
  {\bibfnamefont {K.~T.}\ \bibnamefont {Flanagan}}, \bibinfo {author}
  {\bibfnamefont {C.}~\bibnamefont {Forss{\'e}n}}, \bibinfo {author}
  {\bibfnamefont {S.}~\bibnamefont {Franchoo}}, \bibinfo {author}
  {\bibfnamefont {R.~F.~Garcia}\ \bibnamefont {Ruiz}}, \bibinfo {author}
  {\bibfnamefont {F.~P.}\ \bibnamefont {Gustafsson}}, \bibinfo {author}
  {\bibfnamefont {G.}~\bibnamefont {Hagen}}, \bibinfo {author} {\bibfnamefont
  {G.~R.}\ \bibnamefont {Jansen}}, \bibinfo {author} {\bibfnamefont
  {A.}~\bibnamefont {Kanellakopoulos}}, \bibinfo {author} {\bibfnamefont
  {M.}~\bibnamefont {Kortelainen}}, \bibinfo {author} {\bibfnamefont
  {W.}~\bibnamefont {Nazarewicz}}, \bibinfo {author} {\bibfnamefont
  {G.}~\bibnamefont {Neyens}}, \bibinfo {author} {\bibfnamefont
  {T.}~\bibnamefont {Papenbrock}}, \bibinfo {author} {\bibfnamefont {P.~G.}\
  \bibnamefont {Reinhard}}, \bibinfo {author} {\bibfnamefont {C.~M.}\
  \bibnamefont {Ricketts}}, \bibinfo {author} {\bibfnamefont {B.~K.}\
  \bibnamefont {Sahoo}}, \bibinfo {author} {\bibfnamefont {A.~R.}\ \bibnamefont
  {Vernon}}, \ and\ \bibinfo {author} {\bibfnamefont {S.~G.}\ \bibnamefont
  {Wilkins}},\ }\bibfield  {title} {\enquote {\bibinfo {title} {Charge radii of
  exotic potassium isotopes challenge nuclear theory and the magic character of
  n = 32},}\ }\href {\doibase 10.1038/s41567-020-01136-5} {\bibfield  {journal}
  {\bibinfo  {journal} {Nature Physics}\ } (\bibinfo {year} {2021}),\
  10.1038/s41567-020-01136-5}\BibitemShut {NoStop}%
\bibitem [{\citenamefont {Efros}\ \emph {et~al.}(1994)\citenamefont {Efros},
  \citenamefont {Leidemann},\ and\ \citenamefont {Orlandini}}]{efros1994}%
  \BibitemOpen
  \bibfield  {author} {\bibinfo {author} {\bibfnamefont {Victor~D.}\
  \bibnamefont {Efros}}, \bibinfo {author} {\bibfnamefont {Winfred}\
  \bibnamefont {Leidemann}}, \ and\ \bibinfo {author} {\bibfnamefont
  {Giuseppina}\ \bibnamefont {Orlandini}},\ }\bibfield  {title} {\enquote
  {\bibinfo {title} {Response functions from integral transforms with a lorentz
  kernel},}\ }\href {\doibase 10.1016/0370-2693(94)91355-2} {\bibfield
  {journal} {\bibinfo  {journal} {Phys. Lett. B}\ }\textbf {\bibinfo {volume}
  {338}},\ \bibinfo {pages} {130 -- 133} (\bibinfo {year} {1994})}\BibitemShut
  {NoStop}%
\bibitem [{\citenamefont {Efros}\ \emph {et~al.}(2007)\citenamefont {Efros},
  \citenamefont {Leidemann}, \citenamefont {Orlandini},\ and\ \citenamefont
  {Barnea}}]{efros2007}%
  \BibitemOpen
  \bibfield  {author} {\bibinfo {author} {\bibfnamefont {V~D}\ \bibnamefont
  {Efros}}, \bibinfo {author} {\bibfnamefont {W}~\bibnamefont {Leidemann}},
  \bibinfo {author} {\bibfnamefont {G}~\bibnamefont {Orlandini}}, \ and\
  \bibinfo {author} {\bibfnamefont {N}~\bibnamefont {Barnea}},\ }\bibfield
  {title} {\enquote {\bibinfo {title} {The lorentz integral transform (lit)
  method and its applications to perturbation-induced reactions},}\ }\href
  {http://stacks.iop.org/0954-3899/34/i=12/a=R02} {\bibfield  {journal}
  {\bibinfo  {journal} {Journal of Physics G: Nuclear and Particle Physics}\
  }\textbf {\bibinfo {volume} {34}},\ \bibinfo {pages} {R459} (\bibinfo {year}
  {2007})}\BibitemShut {NoStop}%
\bibitem [{\citenamefont {Bacca}\ \emph {et~al.}(2013)\citenamefont {Bacca},
  \citenamefont {Barnea}, \citenamefont {Hagen}, \citenamefont {Orlandini},\
  and\ \citenamefont {Papenbrock}}]{bacca2013}%
  \BibitemOpen
  \bibfield  {author} {\bibinfo {author} {\bibfnamefont {S.}~\bibnamefont
  {Bacca}}, \bibinfo {author} {\bibfnamefont {N.}~\bibnamefont {Barnea}},
  \bibinfo {author} {\bibfnamefont {G.}~\bibnamefont {Hagen}}, \bibinfo
  {author} {\bibfnamefont {G.}~\bibnamefont {Orlandini}}, \ and\ \bibinfo
  {author} {\bibfnamefont {T.}~\bibnamefont {Papenbrock}},\ }\bibfield  {title}
  {\enquote {\bibinfo {title} {First principles description of the giant dipole
  resonance in $^{16}\mathbf{O}$},}\ }\href {\doibase
  10.1103/PhysRevLett.111.122502} {\bibfield  {journal} {\bibinfo  {journal}
  {Phys. Rev. Lett.}\ }\textbf {\bibinfo {volume} {111}},\ \bibinfo {pages}
  {122502} (\bibinfo {year} {2013})}\BibitemShut {NoStop}%
\bibitem [{\citenamefont {{Bacca}}\ \emph {et~al.}(2014)\citenamefont
  {{Bacca}}, \citenamefont {{Barnea}}, \citenamefont {{Hagen}}, \citenamefont
  {{Miorelli}}, \citenamefont {{Orlandini}},\ and\ \citenamefont
  {{Papenbrock}}}]{bacca2014}%
  \BibitemOpen
  \bibfield  {author} {\bibinfo {author} {\bibfnamefont {S.}~\bibnamefont
  {{Bacca}}}, \bibinfo {author} {\bibfnamefont {N.}~\bibnamefont {{Barnea}}},
  \bibinfo {author} {\bibfnamefont {G.}~\bibnamefont {{Hagen}}}, \bibinfo
  {author} {\bibfnamefont {M.}~\bibnamefont {{Miorelli}}}, \bibinfo {author}
  {\bibfnamefont {G.}~\bibnamefont {{Orlandini}}}, \ and\ \bibinfo {author}
  {\bibfnamefont {T.}~\bibnamefont {{Papenbrock}}},\ }\bibfield  {title}
  {\enquote {\bibinfo {title} {{Giant and pigmy dipole resonances in 4He,
  16,22O, and 40Ca from chiral nucleon-nucleon interactions}},}\ }\href
  {http://adsabs.harvard.edu/abs/2014arXiv1410.2258B} {\bibfield  {journal}
  {\bibinfo  {journal} {ArXiv e-prints}\ } (\bibinfo {year} {2014})},\ \Eprint
  {http://arxiv.org/abs/1410.2258} {arXiv:1410.2258 [nucl-th]} \BibitemShut
  {NoStop}%
\bibitem [{\citenamefont {Sobczyk}\ \emph {et~al.}(2020)\citenamefont
  {Sobczyk}, \citenamefont {Acharya}, \citenamefont {Bacca},\ and\
  \citenamefont {Hagen}}]{Sobczyk:2020qtw}%
  \BibitemOpen
  \bibfield  {author} {\bibinfo {author} {\bibfnamefont {J.~E.}\ \bibnamefont
  {Sobczyk}}, \bibinfo {author} {\bibfnamefont {B.}~\bibnamefont {Acharya}},
  \bibinfo {author} {\bibfnamefont {S.}~\bibnamefont {Bacca}}, \ and\ \bibinfo
  {author} {\bibfnamefont {G.}~\bibnamefont {Hagen}},\ }\bibfield  {title}
  {\enquote {\bibinfo {title} {{Coulomb sum rule for $^4$He and $^{16}$O from
  coupled-cluster theory}},}\ }\href {\doibase 10.1103/PhysRevC.102.064312}
  {\bibfield  {journal} {\bibinfo  {journal} {Phys. Rev. C}\ }\textbf {\bibinfo
  {volume} {102}},\ \bibinfo {pages} {064312} (\bibinfo {year} {2020})},\
  \Eprint {http://arxiv.org/abs/2009.01761} {arXiv:2009.01761 [nucl-th]}
  \BibitemShut {NoStop}%
\bibitem [{\citenamefont {Ekstr\"om}\ \emph {et~al.}(2015)\citenamefont
  {Ekstr\"om}, \citenamefont {Jansen}, \citenamefont {Wendt}, \citenamefont
  {Hagen}, \citenamefont {Papenbrock}, \citenamefont {Carlsson}, \citenamefont
  {Forss\'en}, \citenamefont {Hjorth-Jensen}, \citenamefont {Navr\'atil},\ and\
  \citenamefont {Nazarewicz}}]{Ekstrom:2015rta}%
  \BibitemOpen
  \bibfield  {author} {\bibinfo {author} {\bibfnamefont {A.}~\bibnamefont
  {Ekstr\"om}}, \bibinfo {author} {\bibfnamefont {G.~R.}\ \bibnamefont
  {Jansen}}, \bibinfo {author} {\bibfnamefont {K.~A.}\ \bibnamefont {Wendt}},
  \bibinfo {author} {\bibfnamefont {G.}~\bibnamefont {Hagen}}, \bibinfo
  {author} {\bibfnamefont {T.}~\bibnamefont {Papenbrock}}, \bibinfo {author}
  {\bibfnamefont {B.~D.}\ \bibnamefont {Carlsson}}, \bibinfo {author}
  {\bibfnamefont {C.}~\bibnamefont {Forss\'en}}, \bibinfo {author}
  {\bibfnamefont {M.}~\bibnamefont {Hjorth-Jensen}}, \bibinfo {author}
  {\bibfnamefont {P.}~\bibnamefont {Navr\'atil}}, \ and\ \bibinfo {author}
  {\bibfnamefont {W.}~\bibnamefont {Nazarewicz}},\ }\bibfield  {title}
  {\enquote {\bibinfo {title} {{Accurate nuclear radii and binding energies
  from a chiral interaction}},}\ }\href {\doibase 10.1103/PhysRevC.91.051301}
  {\bibfield  {journal} {\bibinfo  {journal} {Phys. Rev. C}\ }\textbf {\bibinfo
  {volume} {91}},\ \bibinfo {pages} {051301} (\bibinfo {year} {2015})},\
  \Eprint {http://arxiv.org/abs/1502.04682} {arXiv:1502.04682 [nucl-th]}
  \BibitemShut {NoStop}%
\bibitem [{\citenamefont {Jiang}\ \emph {et~al.}(2020)\citenamefont {Jiang},
  \citenamefont {Ekstr\"om}, \citenamefont {Forss\'en}, \citenamefont {Hagen},
  \citenamefont {Jansen},\ and\ \citenamefont {Papenbrock}}]{jiang2020}%
  \BibitemOpen
  \bibfield  {author} {\bibinfo {author} {\bibfnamefont {W.~G.}\ \bibnamefont
  {Jiang}}, \bibinfo {author} {\bibfnamefont {A.}~\bibnamefont {Ekstr\"om}},
  \bibinfo {author} {\bibfnamefont {C.}~\bibnamefont {Forss\'en}}, \bibinfo
  {author} {\bibfnamefont {G.}~\bibnamefont {Hagen}}, \bibinfo {author}
  {\bibfnamefont {G.~R.}\ \bibnamefont {Jansen}}, \ and\ \bibinfo {author}
  {\bibfnamefont {T.}~\bibnamefont {Papenbrock}},\ }\bibfield  {title}
  {\enquote {\bibinfo {title} {Accurate bulk properties of nuclei from $a=2$ to
  $\ensuremath{\infty}$ from potentials with $\mathrm{\ensuremath{\Delta}}$
  isobars},}\ }\href {\doibase 10.1103/PhysRevC.102.054301} {\bibfield
  {journal} {\bibinfo  {journal} {Phys. Rev. C}\ }\textbf {\bibinfo {volume}
  {102}},\ \bibinfo {pages} {054301} (\bibinfo {year} {2020})}\BibitemShut
  {NoStop}%
\bibitem [{\citenamefont {Kelly}(2004)}]{Kelly:2004hm}%
  \BibitemOpen
  \bibfield  {author} {\bibinfo {author} {\bibfnamefont {J.J.}\ \bibnamefont
  {Kelly}},\ }\bibfield  {title} {\enquote {\bibinfo {title} {{Simple
  parametrization of nucleon form factors}},}\ }\href {\doibase
  10.1103/PhysRevC.70.068202} {\bibfield  {journal} {\bibinfo  {journal} {Phys.
  Rev. C}\ }\textbf {\bibinfo {volume} {70}},\ \bibinfo {pages} {068202}
  (\bibinfo {year} {2004})}\BibitemShut {NoStop}%
\bibitem [{\citenamefont {Krebs}\ \emph {et~al.}(2019)\citenamefont {Krebs},
  \citenamefont {Epelbaum},\ and\ \citenamefont {Mei\ss{}ner}}]{Krebs:2019aka}%
  \BibitemOpen
  \bibfield  {author} {\bibinfo {author} {\bibfnamefont {H.}~\bibnamefont
  {Krebs}}, \bibinfo {author} {\bibfnamefont {E.}~\bibnamefont {Epelbaum}}, \
  and\ \bibinfo {author} {\bibfnamefont {U.~G.}\ \bibnamefont {Mei\ss{}ner}},\
  }\bibfield  {title} {\enquote {\bibinfo {title} {{Nuclear Electromagnetic
  Currents to Fourth Order in Chiral Effective Field Theory}},}\ }\href
  {\doibase 10.1007/s00601-019-1500-5} {\bibfield  {journal} {\bibinfo
  {journal} {Few Body Syst.}\ }\textbf {\bibinfo {volume} {60}},\ \bibinfo
  {pages} {31} (\bibinfo {year} {2019})},\ \Eprint
  {http://arxiv.org/abs/1902.06839} {arXiv:1902.06839 [nucl-th]} \BibitemShut
  {NoStop}%
\bibitem [{\citenamefont {Stanton}\ and\ \citenamefont
  {Bartlett}(1993)}]{stanton1993}%
  \BibitemOpen
  \bibfield  {author} {\bibinfo {author} {\bibfnamefont {John~F.}\ \bibnamefont
  {Stanton}}\ and\ \bibinfo {author} {\bibfnamefont {Rodney~J.}\ \bibnamefont
  {Bartlett}},\ }\bibfield  {title} {\enquote {\bibinfo {title} {The equation
  of motion coupled-cluster method. a systematic biorthogonal approach to
  molecular excitation energies, transition probabilities, and excited state
  properties},}\ }\href {\doibase 10.1063/1.464746} {\bibfield  {journal}
  {\bibinfo  {journal} {J. Chem. Phys.}\ }\textbf {\bibinfo {volume} {98}},\
  \bibinfo {pages} {7029--7039} (\bibinfo {year} {1993})}\BibitemShut {NoStop}%
\bibitem [{\citenamefont {Miorelli}\ \emph {et~al.}(2018)\citenamefont
  {Miorelli}, \citenamefont {Bacca}, \citenamefont {Hagen},\ and\ \citenamefont
  {Papenbrock}}]{miorelli2018}%
  \BibitemOpen
  \bibfield  {author} {\bibinfo {author} {\bibfnamefont {M.}~\bibnamefont
  {Miorelli}}, \bibinfo {author} {\bibfnamefont {S.}~\bibnamefont {Bacca}},
  \bibinfo {author} {\bibfnamefont {G.}~\bibnamefont {Hagen}}, \ and\ \bibinfo
  {author} {\bibfnamefont {T.}~\bibnamefont {Papenbrock}},\ }\bibfield  {title}
  {\enquote {\bibinfo {title} {Computing the dipole polarizability of
  $^{48}\mathrm{Ca}$ with increased precision},}\ }\href {\doibase
  10.1103/PhysRevC.98.014324} {\bibfield  {journal} {\bibinfo  {journal} {Phys.
  Rev. C}\ }\textbf {\bibinfo {volume} {98}},\ \bibinfo {pages} {014324}
  (\bibinfo {year} {2018})}\BibitemShut {NoStop}%
\bibitem [{\citenamefont {Acharya}\ and\ \citenamefont {Bacca}(2020)}]{bijaya}%
  \BibitemOpen
  \bibfield  {author} {\bibinfo {author} {\bibfnamefont {Bijaya}\ \bibnamefont
  {Acharya}}\ and\ \bibinfo {author} {\bibfnamefont {Sonia}\ \bibnamefont
  {Bacca}},\ }\bibfield  {title} {\enquote {\bibinfo {title} {Neutrino-deuteron
  scattering: Uncertainty quantification and new ${L}_{1,A}$ constraints},}\
  }\href {\doibase 10.1103/PhysRevC.101.015505} {\bibfield  {journal} {\bibinfo
   {journal} {Phys. Rev. C}\ }\textbf {\bibinfo {volume} {101}},\ \bibinfo
  {pages} {015505} (\bibinfo {year} {2020})}\BibitemShut {NoStop}%
\bibitem [{\citenamefont {Barnea}\ \emph {et~al.}(2001)\citenamefont {Barnea},
  \citenamefont {Leidemann},\ and\ \citenamefont {Orlandini}}]{barnea2001}%
  \BibitemOpen
  \bibfield  {author} {\bibinfo {author} {\bibfnamefont {Nir}\ \bibnamefont
  {Barnea}}, \bibinfo {author} {\bibfnamefont {Winfried}\ \bibnamefont
  {Leidemann}}, \ and\ \bibinfo {author} {\bibfnamefont {Giuseppina}\
  \bibnamefont {Orlandini}},\ }\bibfield  {title} {\enquote {\bibinfo {title}
  {State-dependent effective interaction for the hyperspherical formalism with
  noncentral forces},}\ }\href {\doibase
  http://dx.doi.org/10.1016/S0375-9474(01)00794-1} {\bibfield  {journal}
  {\bibinfo  {journal} {Nuclear Physics A}\ }\textbf {\bibinfo {volume}
  {693}},\ \bibinfo {pages} {565 -- 578} (\bibinfo {year} {2001})}\BibitemShut
  {NoStop}%
\bibitem [{\citenamefont {Rocco}\ \emph {et~al.}(2018)\citenamefont {Rocco},
  \citenamefont {Leidemann}, \citenamefont {Lovato},\ and\ \citenamefont
  {Orlandini}}]{Rocco:2018tes}%
  \BibitemOpen
  \bibfield  {author} {\bibinfo {author} {\bibfnamefont {Noemi}\ \bibnamefont
  {Rocco}}, \bibinfo {author} {\bibfnamefont {Winfried}\ \bibnamefont
  {Leidemann}}, \bibinfo {author} {\bibfnamefont {Alessandro}\ \bibnamefont
  {Lovato}}, \ and\ \bibinfo {author} {\bibfnamefont {Giuseppina}\ \bibnamefont
  {Orlandini}},\ }\bibfield  {title} {\enquote {\bibinfo {title} {{Relativistic
  effects in ab-initio electron-nucleus scattering}},}\ }\href {\doibase
  10.1103/PhysRevC.97.055501} {\bibfield  {journal} {\bibinfo  {journal} {Phys.
  Rev. C}\ }\textbf {\bibinfo {volume} {97}},\ \bibinfo {pages} {055501}
  (\bibinfo {year} {2018})},\ \Eprint {http://arxiv.org/abs/1801.07111}
  {arXiv:1801.07111 [nucl-th]} \BibitemShut {NoStop}%
\bibitem [{\citenamefont {Carlson}\ \emph {et~al.}(2002)\citenamefont
  {Carlson}, \citenamefont {Jourdan}, \citenamefont {Schiavilla},\ and\
  \citenamefont {Sick}}]{world_data}%
  \BibitemOpen
  \bibfield  {author} {\bibinfo {author} {\bibfnamefont {J.}~\bibnamefont
  {Carlson}}, \bibinfo {author} {\bibfnamefont {J.}~\bibnamefont {Jourdan}},
  \bibinfo {author} {\bibfnamefont {R.}~\bibnamefont {Schiavilla}}, \ and\
  \bibinfo {author} {\bibfnamefont {I.}~\bibnamefont {Sick}},\ }\bibfield
  {title} {\enquote {\bibinfo {title} {Longitudinal and transverse quasielastic
  response functions of light nuclei},}\ }\href {\doibase
  10.1103/PhysRevC.65.024002} {\bibfield  {journal} {\bibinfo  {journal} {Phys.
  Rev. C}\ }\textbf {\bibinfo {volume} {65}},\ \bibinfo {pages} {024002}
  (\bibinfo {year} {2002})}\BibitemShut {NoStop}%
\bibitem [{\citenamefont {Bacca}\ \emph {et~al.}(2009)\citenamefont {Bacca},
  \citenamefont {Barnea}, \citenamefont {Leidemann},\ and\ \citenamefont
  {Orlandini}}]{BaccaPRL2009}%
  \BibitemOpen
  \bibfield  {author} {\bibinfo {author} {\bibfnamefont {Sonia}\ \bibnamefont
  {Bacca}}, \bibinfo {author} {\bibfnamefont {Nir}\ \bibnamefont {Barnea}},
  \bibinfo {author} {\bibfnamefont {Winfried}\ \bibnamefont {Leidemann}}, \
  and\ \bibinfo {author} {\bibfnamefont {Giuseppina}\ \bibnamefont
  {Orlandini}},\ }\bibfield  {title} {\enquote {\bibinfo {title} {Role of the
  final-state interaction and three-body force on the longitudinal response
  function of $^{4}\mathrm{He}$},}\ }\href {\doibase
  10.1103/PhysRevLett.102.162501} {\bibfield  {journal} {\bibinfo  {journal}
  {Phys. Rev. Lett.}\ }\textbf {\bibinfo {volume} {102}},\ \bibinfo {pages}
  {162501} (\bibinfo {year} {2009})}\BibitemShut {NoStop}%
\bibitem [{\citenamefont {Lonardoni}\ \emph {et~al.}(2017)\citenamefont
  {Lonardoni}, \citenamefont {Lovato}, \citenamefont {Pieper},\ and\
  \citenamefont {Wiringa}}]{Lonardoni}%
  \BibitemOpen
  \bibfield  {author} {\bibinfo {author} {\bibfnamefont {D.}~\bibnamefont
  {Lonardoni}}, \bibinfo {author} {\bibfnamefont {A.}~\bibnamefont {Lovato}},
  \bibinfo {author} {\bibfnamefont {Steven~C.}\ \bibnamefont {Pieper}}, \ and\
  \bibinfo {author} {\bibfnamefont {R.~B.}\ \bibnamefont {Wiringa}},\
  }\bibfield  {title} {\enquote {\bibinfo {title} {Variational calculation of
  the ground state of closed-shell nuclei up to $a=40$},}\ }\href {\doibase
  10.1103/PhysRevC.96.024326} {\bibfield  {journal} {\bibinfo  {journal} {Phys.
  Rev. C}\ }\textbf {\bibinfo {volume} {96}},\ \bibinfo {pages} {024326}
  (\bibinfo {year} {2017})}\BibitemShut {NoStop}%
\bibitem [{\citenamefont {Williamson}\ \emph {et~al.}(1997)\citenamefont
  {Williamson} \emph {et~al.}}]{Williamson:1997zz}%
  \BibitemOpen
  \bibfield  {author} {\bibinfo {author} {\bibfnamefont {C.~F.}\ \bibnamefont
  {Williamson}} \emph {et~al.},\ }\bibfield  {title} {\enquote {\bibinfo
  {title} {{Quasielastic electron scattering from Ca-40}},}\ }\href {\doibase
  10.1103/PhysRevC.56.3152} {\bibfield  {journal} {\bibinfo  {journal} {Phys.
  Rev. C}\ }\textbf {\bibinfo {volume} {56}},\ \bibinfo {pages} {3152--3172}
  (\bibinfo {year} {1997})}\BibitemShut {NoStop}%
\bibitem [{com()}]{com_mom_distr}%
  \BibitemOpen
  \href@noop {} {}\bibinfo {howpublished} {J.~E.~Sobczyk, B.~Acharya, S.~Bacca,
  G.~Hagen, and T.~Papenbrock, in preparation.}\BibitemShut {Stop}%
\bibitem [{Dor()}]{Doria}%
  \BibitemOpen
  \href@noop {} {}\bibinfo {howpublished} {L.~Doria and M. Mihovilovic, private
  communication.}\BibitemShut {Stop}%
\end{thebibliography}%

\end{document}